\documentclass[12pt,a4paper]{article}

\usepackage{amsmath}
\usepackage{amsfonts}
\usepackage{amssymb}
\usepackage{graphicx}
\usepackage{array}
\usepackage{cancel}
\usepackage{caption}
\usepackage{wrapfig}
\usepackage{mathrsfs}
\usepackage{cancel}
\usepackage{siunitx}
\usepackage{amsthm}
\usepackage{tikz-cd}
\usetikzlibrary{babel}
\usepackage[colorlinks=true,
linkcolor=blue,
urlcolor=blue,citecolor=green]{hyperref}
\usepackage[left=1.5cm,right=1.5cm,top=1.5cm,bottom=1.5cm,includefoot,footskip=1.5cm]{geometry}
\usepackage[OT1, T2A]{fontenc}
\usepackage[utf8]{inputenc}
\usepackage[english]{babel}
\usepackage{csquotes}

\usepackage[sorting=none,
backend=biber
]{biblatex}
\usepackage[toc]{appendix}

\let\{=\lbrace
\let\}=\rbrace
\let\[=\lbracket
\let\]=\rbracket
\let\dg=\dagger

\let\lc = \varepsilon
\let\ep = \epsilon
\let\w = \omega

\def\ili{\int\limits}
\def\<{\left(}
\def\>{\right)}
\def\bd{\boldsymbol}

\def\l|{\left|}
\def\r|{\right|}

\def\p{\partial}

\def\const{\text{const}}

\def\grg{\mathfrak{g}}

\newcommand{\pdf}[2]{\dfrac{\partial #1}{\partial #2}}

\newcommand{\braket}[2]{\langle #1|#2 \rangle}
\newcommand{\bl}{\left\langle}
\newcommand{\br}{\right\rangle}

\newcommand{\mt}[1]{\mathcal{#1}}

\DeclareMathOperator{\Tr}{Tr}

\DeclareMathOperator{\sign}{sign}

\numberwithin{equation}{section}
\usepackage{authblk}
\author[1,2]{E.T.Akhmedov}
\author[1,3,4]{A.A.Artemev}
\author[1,3,4]{I.V.Kochergin}
\affil[1]{Moscow Institute of Physics and Technology, Institutskii per. 9, 141700, Dolgoprudny, Russia}
\affil[2]{ Institute for Theoretical and Experimental Physics, B. Cheremushkinskaya 25, 117218, Moscow, Russia}
\affil[3]{Landau Institute for Theoretical Physics, 142432, Chernogolovka, Russia}
\affil[4]{Skolkovo Institute of Science and Technology, 121205, Moscow, Russia}
\title{\textcolor{black}{On interacting quantum fields in various charts of anti de Sitter space--time}}

\begin{document}

\maketitle

\begin{abstract}

The goal of the present note is to understand whether it is possible to define interacting quantum field theory in global anti de Sitter space--time with Lorentzian signature, in its covering space--time (whose time coordinate is not periodic) and in its Poincare patch. We show that in global anti de Sitter space--time there are certain problems to define quantum field theory properly. This is due to an additional UV singularity of the Feynman propagator which is sitting on the light--cone emanating from the antipodal point of the source. There is no such singularity in flat space--time. At the same time quantum field theory in Poincare region of the AdS space--time can be well defined and is related to the one in Euclidean AdS via the analytical continuation. In principle one can also define and analytically continue quantum field theory in the covering anti de Sitter space--time. However, to do that one has to use an unusual $i\epsilon$--prescription in the Feynman propagator, which cannot be used in loop calculations in non--stationary situations. 

\end{abstract}
\newpage

\section{Introduction}

Classical and quantum field theory in anti de Sitter (AdS) space is quite well developed: see \cite{Maldacena1999largeN} and \cite{Moschella:2006pkh} for reviews and e.g.  \cite{Bros:2001tk}, \cite{Bros:2011vh}, \cite{Moschella:2007zza} for the approach close to the one adopted in the present note. Loop corrections in AdS have been considered in many places: see e.g. \cite{Akhmedov:2012hk}, \cite{Bertan:2018khc}, \cite{Carmi:2019ocp}, \cite{Ponomarev:2019ltz}, \cite{Bertan:2018afl} \cite{Sleight:2017cax}, \cite{Giombi:2017hpr} for an incomplete list of references. 

However, to the best of our knowledge in all of these cases quantum loop corrections have been calculated in AdS space with Euclidean signature (EAdS). Meanwhile, loop corrections in Lorentzian AdS space have some peculiarities \cite{Akhmedov2018Ultraviolet} related to the presence of additional UV singularities in the propagators, which are sitting on the light--cone emanating from the antipodal point of the source. Such singularities are not present in flat space--time and in EAdS.

The goal of the present note is to understand whether it is possible to define interacting quantum field theory in global AdS space--time with Lorentzian signature, in its covering space--time (CAdS) and in its Poincare patch (PP). 

Namely, in this note we show that both in global AdS and in CAdS there are certain problems to define quantum field theory properly. That is exactly due to the additional UV singularity of the propagator. Quantum field theory in PP of the AdS space--time can be well defined and is related to the one in EAdS via the analytical continuation. One also can define quantum field theory in the CAdS, but with the use of a peculiar $i\epsilon$--shift of the UV singularities of the Feynman propagator in the coordinate space.

The paper is organized as follows. In section \ref{adsint} we consider classical focusing of geodesics in global AdS space--time and study how generic it is for asymptotically AdS spaces in two and higher dimensions. We would like to understand if there is a general relation of this focusing to the presence of the additional (non--local) singularity in the Feynman propagator. In section \ref{nlsing} we study in detail the singular behavior of the two-point Wightman function of the free scalar field theory in two--dimensional asymptotically AdS spaces. We also consider higher dimensions and discuss the relation of the extra singularity to the geodesic focusing for generic space--times of a certain class. The problems associated with perturbation theory and possible approaches to loop calculations in global AdS space--time are discussed in section \ref{pert}. Section \ref{cvpnk} is dedicated to the isometry invariance of correlation functions in interacting theories in the CAdS space--time and in the PP. We also study the analytical continuation in quantum field theories in CAds and PP to the Euclidean AdS. To make the paper self-contained we provide some technical details in the appendices.

\section{Geodesics in asymptotically anti de Sitter space--times}\label{adsint}

\noindent We start our discussion with the classical focusing of geodesics, which may be related to the presence of anomalous non--local UV singularities in propagators in global AdS space--time. In this note we refer to the standard UV singularity of the propagator as local: it is located on the light cone emanating from the source --- it becomes truly local after the analytical continuation to the Euclidean signature. We call the UV singularity as non--local when it appears on the light cone emanating from the antipodal point of the source.

The $d$-dimensional AdS space with the Lorentzian signature is the hyperboloid in the $(d+1)$--dimensional flat embedding space--time. We adopt the signature of the metric of the embedding space as follows $(-,\dots,-,+,+)$. The equation for the hyperboloid is:

\begin{equation}
X_A X^A = R^2,
\end{equation}
where $X^A$, $I = (0,\dots,d)$ are coordinates in the embedding flat space--time. To simplify equations below we set $R = 1$. One can introduce the global coordinates parametrizing the entire hyperboloid:

\begin{equation}
X^i = \psi^i\tan\theta,\quad X^{d-1} = \dfrac{\cos\tau}{\cos\theta},\quad X^d = \dfrac{\sin\tau}{\cos\theta},\label{embcoord}
\end{equation}
where $i = 1,\dots, d-2$ and $\psi^i$ are coordinates on $(d-2)$--dimensional sphere of unit radius: $\sum_i (\psi^i)^2 = 1$. The map is single-valued if $\tau \in [0,2\pi)$ and $\theta \in [0,\pi/2]$ for $d > 2$ or $\theta \in [-\pi/2,\pi/2]$ for $d=2$. 

 The AdS metric in terms of these coordinates is as follows:
\begin{equation}
ds^2_{\text{AdS}_d} = \dfrac{d\tau^2-d\theta^2-\sin^2\theta\,d\Omega^2_{d-2}}{\cos^2\theta} = \dfrac{d\tau^2 - d\Omega_{d-1}^2}{\cos^2\theta},\label{adsmetr}
\end{equation}
where $d\Omega_{d-1}^2$ is the  metric on $(d-1)$--dimensional sphere. This metric is conformally equivalent to the one of the Einstein Static Universe (ESU)~--- the $S^1 \times S^{d-1}$  space. However, in ESU the range of the $\theta$ coordinate is as follows $\theta \in [0,\pi]$ for $d > 2$ and $\theta \in (-\pi,\pi]$ for $d = 2$. 

Thus, the AdS space can be Weyl--mapped only onto the half of ESU. It means that to define quantum field theory on the AdS space--time one has to impose boundary conditions at $\theta = \pi/2$ (or at $\theta = \pm \pi/2$ for $d=2$). In fact, the lightlike geodesics can approach this boundary in finite proper time of the internal AdS observer. This problem was studied in \cite{Avis1978AdS}. Its brief review for the $d=2$ case is given in Appendix \ref{massl}. In summary, we impose the reflective boundary conditions, i.e. assume that the modes are zero at the boundary. We will do the same also in asymptotically AdS space--times, which will be defined below.

We denote the embedding map (\ref{embcoord}) by $X^A = f^A(x)$, where $x^\mu$ are the coordinates in the AdS space--time. The AdS isometry invariant,

\begin{equation}
\zeta(x,x') = f_A(x) f^A(x') = \dfrac{\cos(\tau-\tau') - \sin\theta\sin\theta' \psi^i \psi'^i}{\cos\theta\cos\theta'},\label{gpar}
\end{equation} 
is related to the geodesic distance between two points $x$ and $x'$ of the AdS space $d(x,x')$. Namely, $\cos d(x,x') = \zeta(x,x')$. The separation between two points is timelike if $|\zeta|<1$ and spacelike if $\zeta > 1$. If $\zeta < -1$ both real and imaginary parts of $d$ are nonzero and there are no geodesics connecting $x$ and $x'$. For convenience we will call the separation between two points spacelike if they cannot be connected by a timelike geodesics

The global AdS manifold is not globally hyperbolic for two reasons. First, due to the presence of the boundary that we have discussed above and due to the presence of closed timelike geodesics, because $\tau \in [0,2\pi)$. Second, in global AdS space all geodesics originated from a point $x$ intersect again at its antipodal point, which we denote as $\bar{x}$ ($\bar{X} = -X$ if $X = f(x)$):

\begin{equation}
\bar{\tau} = \tau+\pi,\quad \bar{\psi} = -\psi,\quad \bar{\theta} = \begin{cases}
\theta,& d>2\\
-\theta,&d=2.
\end{cases}
\end{equation}
This fact is related to the necessity to impose the aforementioned boundary conditions for the massless fields.

In the covering global AdS manifold (CAdS), which has the same metric as (\ref{adsmetr}), but with $\tau \in (-\infty, +\infty)$, the problem of the presence of the closed timelike geodesics is resolved. But the manifold is still not globally hyperbolic due to the necessity to impose boundary conditions on top of initial ones. Due to the $2\pi$-periodicity of geodesics the condition $|\zeta|<1$ of timelike separation still holds.

Also in the CAdS space--time one encounters the phenomenon of the focusing of timelike geodesics, which is the remnant of the presence of closed timelike curves in global AdS. In this note we discuss this phenomenon from different perspectives on classical and quantum level. The main goal of this study is to understand the interrelation between the presence of such a focusing and the presence of the extra UV singularity of the propagators in global AdS and CAdS space--times.



Another frequently used coordinate chart in AdS space is the one where the metric is conformally flat. It covers only a half of AdS space~--- the PP $X^{d-1} > X^0$:

\begin{equation}
X^0 = \dfrac{z^2 -1+ \bd{x}^2-t^2}{2z},\quad X^i = \dfrac{x^i}{z},\quad X^{d-1} = \dfrac{z^2+1 + \bd{x}^2-t^2}{2z},\quad X^d = \dfrac{t}{z},\label{poincare}
\end{equation}
where $z>0$. The metric and the isometry invariant (which is related to the geodesic distance) of this chart are as follows:

\begin{equation}
ds^2 = \dfrac{dt^2-dz^2 - \sum_i (dx^i)^2}{z^2},\quad \zeta(x,x') = \dfrac{(z^2+z'^2) + \<\bd{x}-\bd{x}'\>^2 - (t-t')^2}{2zz'}.\label{PoincareMetr}
\end{equation}
Note that the antipod of any point in the PP lies outside of the patch, unless the point is sitting on the boundary of the PP. In the latter case the antipodal point is also sitting on the boundary.


\subsection{Classical geodesic focusing in asymptotically AdS space--times}\label{GFoc}

\noindent In this section we want to understand the physical origin of the classical focusing of geodesics and its relation to the presence of the non--local UV singularity in the propagators for general space--times of a certain class, which is specified below. For simplicity we consider a two-dimensional Lorentzian manifold which can be Weyl--mapped into the half of ESU~--- the asymptotically AdS space--time. We adopt the terminology of \cite{Maldacena1999largeN}.

It is convenient to move to the CAdS space--time by unfolding the time coordinate. The general metric of interest for us is as follows:

\begin{equation}
ds^2 = f(\theta)(d\tau^2 - d\theta^2),\quad \tau \in (-\infty,+\infty),\quad \theta \in (-\pi/2,\pi/2).\label{genmetr}
\end{equation}
We assume that the function $f(\theta)$ is positive and symmetric $f(\theta) = f(-\theta)$ (as we will see, this condition is necessary for geodesics to intercept in the antipodal point) and has power-like singularities when $\theta$ approaches the boundary $\pm \pi/2$:

\begin{equation}
f(\theta) \approx \dfrac{C}{(\pi/2-\theta)^\alpha},\quad \theta \to \pm \frac{\pi}{2},\quad C >0,\label{f}
\end{equation}
where $\alpha > 0$. The latter condition is not essential for the classical consideration, but it will be used in quantum problem in the section that follows below. 

The geodesics can be found via the solution of the Hamilton-Jacobi equation:
\begin{equation}
\<\pdf{\mathbb{S}}{\tau}\>^2 - \<\pdf{\mathbb{S}}{\theta}\>^2 = m^2f(\theta),
\end{equation}
here $\mathbb{S}$ is the minimum single particle action, $m$ is its mass. This mass can be absorbed into $f(\theta)$ (so we can set $m=1$), in pure AdS case we have $f_{\text{AdS}}(\theta) = \frac{m^2}{\cos^2\theta}$. In such a case it means that in \eqref{f} $C=m^2$.

The variables can be separated by the standard substitution $\mathbb{S} = -\w\tau + S_\theta$, $\w > 0$, then

\begin{equation}
S_\theta = \ili_{\theta_0}^{\theta}\sqrt{\w^2-f(\theta_1)}d\theta_1.\label{Sth}
\end{equation}
For a given $\w$ we have two geodesics corresponding to two directions emanating from $\theta_0$ --- one for $\theta$ starting to change in the direction $\theta > \theta_0$. Then such a geodesic can reflect before the boundary and $\theta$ may become less than $\theta_0$ in the course of time evolution. While the other type of geodesics is obtained when $\theta$ is starting to change in the opposite direction $\theta < \theta_0$. 

In (\ref{Sth}) we have a family of solutions parametrized by $\w$. The geodesic equation is given by $-\frac{\p \mathbb{S}}{\p \w} = \tau_0$ for some constant $\tau_0$:
\begin{equation}
\tau = \tau_0 + \pdf{S_\theta}{\w},\label{taug}
\end{equation}
i.e. the initial point of the geodesic is $(\tau_0,\theta_0)$. 

There are two turning points $\pm \theta_t$ (the points of reflection near the boundary): $f(\theta_t) = \w^2$, $\theta_t > 0$, and we assume that the integration domain in (\ref{Sth}) can contain some number of turns in $\pm\theta_t$.  Due to the symmetry of $f(\theta) = f(-\theta)$ the geodesics under consideration intersect after each half-period at $\theta = \pm\theta_0$. If the action over one period between the turning points is $S$:

\begin{equation}
S = \oint \sqrt{\w^2 - f(\theta)}d\theta,\label{contS}
\end{equation}
then for intersection points $S_\theta = \frac{n}{2}S$, where $n$ is a number of half-periods. As it follows from (\ref{taug}), these points are focal for all geodesics (for all $\omega$) originated in $(\theta_0,\tau_0)$ only if 

\begin{equation}
\pdf{S}{\w} = \text{const}.\label{foccond}
\end{equation}
For large $\w$ (i.e. when $\w^2 \gg C$) the minimal action can be estimated. The turning points can be found from (\ref{f}):

\begin{equation}
\dfrac{\pi}{2}-\theta_t \approx \dfrac{C^{\frac{1}{\alpha}}}{\w^{\frac{2}{\alpha}}} \to 0,\quad \w \to \infty.\label{turn}
\end{equation}
Therefore, in this limit we can approximate the space--time geometry by the infinite rectangular well with boundaries at $\theta = \pm \frac{\pi}{2}$. This can be adopted as the zeroth-order approximation of $f(\theta)$. Then $S \approx 2\pi \w$, as $\w^2 \gg C \sim m^2$ and intersection points are as follows:

\begin{equation}
\tau_n = \tau_0 + n\pi,~\theta_n = (-1)^n\theta_0.
\end{equation}
The condition (\ref{foccond}) is always satisfied asymptotically (when $\w \to \infty$), so we may refer to these points as asymptotic focal. If true focal points do exist, they have to coincide with the asymptotic ones. Hence, from (\ref{taug}) the condition of their existence is as follows:
\begin{equation}
S = 2\pi\w + \delta,\label{foccond2}
\end{equation}
where $\delta$ is a constant. This relation is definitely true in $\text{AdS}_2$. In the latter situation $(\tau_1,\theta_1) = (\bar{\tau}_0,\bar{\theta}_0)$ and $\delta = -2\pi m$, where $m$ is the mass. This can be shown by the direct calculation of the integral in \eqref{contS} when $f(\theta) = \frac{m^2}{\cos^2\theta}$.

Finally, let us show that $f(\theta) = \frac{m^2}{\cos^2\theta}$ is the only symmetric potential with the focusing property within the class \eqref{f} under consideration\footnote{As will be shown in the section that follows, interestingly enough the second singularity in the propagator can appear for a wider class of potentials.}. To do that we will use the standard method of the recovering of the potential from the period of oscillations. 

The focusing condition (\ref{foccond}) can be written in the following form:

\begin{equation}
T(\omega^2) \equiv \dfrac{1}{\w}\pdf{S}{\w} = \oint \dfrac{1}{\sqrt{\w^2-f(\theta)}}d\theta = \dfrac{2\pi}{\w}. 
\end{equation}   
The integral on the LHS is a period of oscillations $T(E)$ of a non-relativistic particle with mass $2$ and energy $E=\w^2$ in the potential $f(\theta)$ \cite{Landau1976Mechanics}. If the potential is symmetric, has only one minimum at $\theta = 0$ and is zero at this point, it can be uniquely deduced from $T(E)$. Let $f(0) = m^2$, then $g(\theta) = f(\theta) - m^2$ satisfies these conditions. \textcolor{black}{The period in such a potential is $T_g(E) = T(E + m^2)$. The inverse function $\theta(g)$ for $\theta > 0$ can be expressed as follows:
\begin{equation}
\theta(g) = \dfrac{1}{4\pi} \ili_0^{g} \dfrac{T_g(E)dE}{\sqrt{g-E}}  = \dfrac{1}{2} \ili_0^{g} \dfrac{dE}{\sqrt{(g-E)(E+m^2)}} =  \dfrac{1}{2}\arccos \dfrac{m^2-g}{m^2+g}.
\end{equation}
Hence, we obtain:
\begin{equation}
f(\theta) = m^2+g(\theta) = \dfrac{m^2}{\cos^2\theta} = f_{\text{AdS}}(\theta).
\end{equation}
However, without the symmetry condition the solution is not unique. The function $\theta(f)$ is double-valued, so let $\theta_1(f)$ denote the part with $\theta \le 0$, and $\theta_2(f)$~--- the part with $\theta \ge 0$. Then
\begin{equation}
\theta_1(f) = -\arccos \dfrac{m}{\sqrt{f}} + h(f),\quad \theta_2(f) = \arccos \dfrac{m}{\sqrt{f}} + h(f),
\end{equation}
where $h(f)$ is such a function that $\lim_{f\to + \infty} h(f) = 0$. If the function $h(f)$ approaches zero slower that $\frac{1}{\sqrt{f}}$, then for $f$ sufficiently close to $+\infty$ there exist a solution with either $\theta_2 > \pi/2$ or $\theta_1 < -\pi/2$. It appears because $\arccos \frac{m}{\sqrt{f}} = \pi/2 - \frac{m}{\sqrt{f}} + o(1/\sqrt{f})$. We consider solutions which do not cross the boundaries at $\theta = \pm \pi/2$; hence, $h(f)$ should decay faster that $\frac{1}{\sqrt{f}}$ and in the leading order the singularities of the potential at the boundaries are determined by the $\arccos$-terms and hence are still quadratic. 
}

\section{The non-local UV singularity of the propagator}\label{nlsing}

\noindent In this section we study the behavior of two-point correlation functions of massless and massive free scalar field theories on global AdS manifold, CAdS and asymptotically AdS space--times. Our goal is to understand the physical origin of another UV singularity in generic space--times and its relation to the geodesic focusing. We mostly concentrate on the two--dimensional case, but at the end of this section we extend our considerations to any dimension. The coordinates are the same as in the previous sections. 

\begin{figure}
\centering\includegraphics[width=0.45\textwidth]{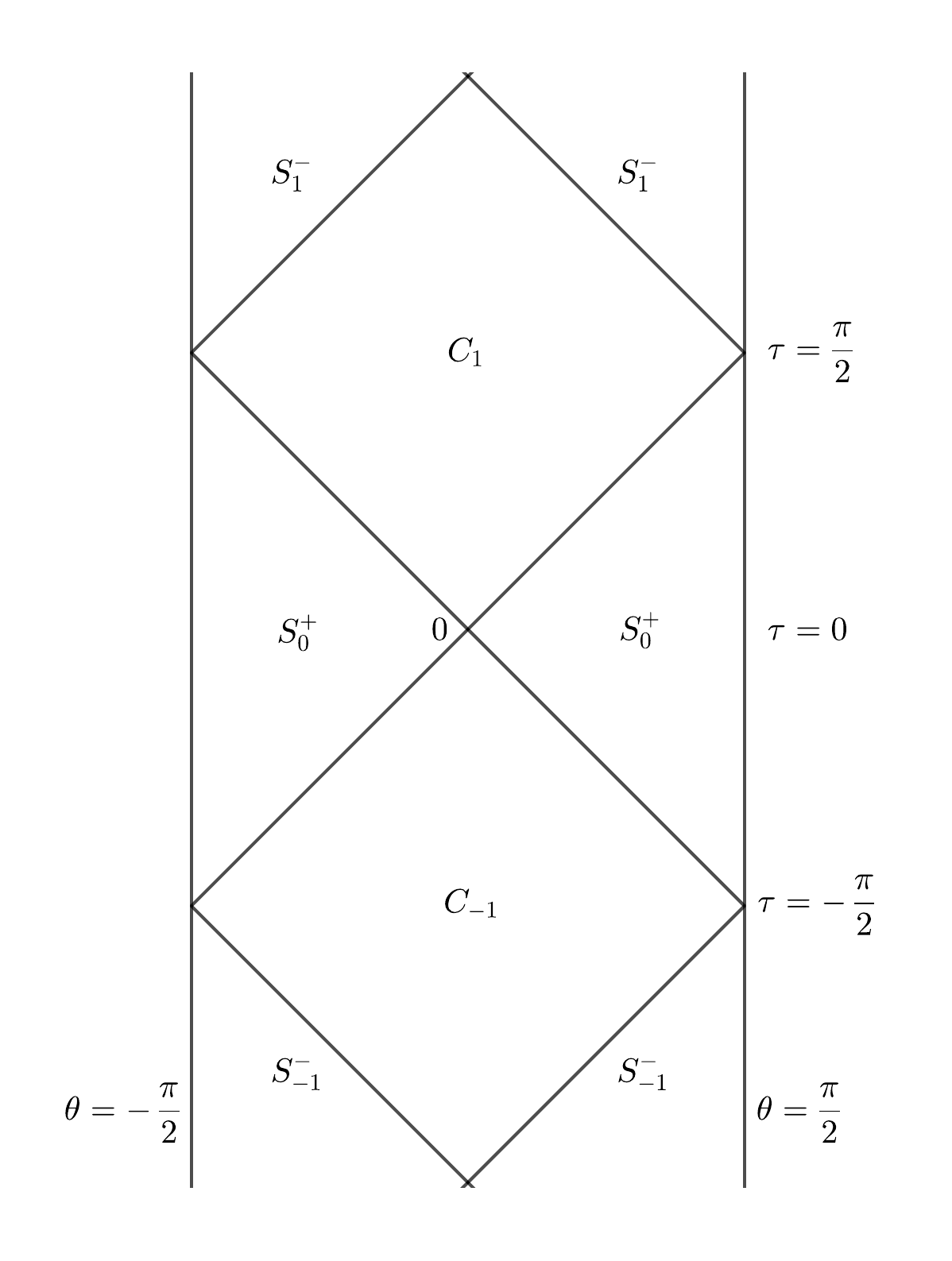}
\caption{The Penrose diagram of the covering AdS} \label{Penrose}
\end{figure}

For the beginning let us point out a few important features of the action of the isometry group in global AdS and CAdS. The isometries of AdS can be extended to the CAdS. Namely, it has the same algebra of Killing vectors (see more details in appendix \ref{massl}) as AdS manifold. Hence, the isometry group (more precisely its connected component of unity) is an exponential of this algebra. We will now show that the relative position of two points in the CAdS cannot change significantly under the action of this group. Let us consider a point $x$ in the CAdS space--time and the set $\mt{X}(x) = \{x': |\zeta(x,x')| = 1\}$. It consists of light cones emanating from $x$, $\bar{x}$ and the images of these points under time translations $\tau \to \tau + 2\pi k$, $k\in \mathbb{Z}$. This set divides the CAdS space--time into regions with $\zeta > 1$ ($S^+_n$ series), $\zeta < -1$ ($S^{-}_n$ series) and $|\zeta| < 1$ ($C_n$ series, such points can be connected to $x$ by a timelike geodesic). The enumeration rule is demonstrated on the fig. \ref{Penrose} for the case $d = 2$, $x = 0$~--- the order of $S^\pm_n$ increases by 1, when $t$ increases by $2\pi$; while the order of $ C_n$ is increased by 1, when $t$ increases by $\pi$. And there is no $C_0$ and $S^-_0$. Generalization to higher dimensions and arbitrary $x$ is straightforward, although in $2d$-case each of $S^{+}_n$ and $S^{-}_n$ in fact consists of two disconnected regions. The isometries transform light cones into light--cones, hence $\mt{X}(gx) = g \mt{X}(x)$ for an isometry $g$. As we consider only the connected subgroup, this isometry preserves the order of regions: $C_n \to C_n$, $S^{\pm}_n \to S^{\pm}_n$ under the action of $g$. This is an important difference of the situation in CAdS with respect to the one in global AdS. We say that $\mt{A}$ (which can be either $S^\pm_n$ or $C_n$) is a relative region of $x$ and $x'$ if $x \in \mt{A}(x')$.

Now let us continue with the properties of the correlation functions.
The Wightman function $W(x,x') \equiv \langle \phi(x) \, \phi(x')\rangle $, say of the real scalar field, in global AdS space--time has an additional UV singularity at the antipodal point\footnote{The regular UV singularity of the propagator is located on the light cone emanating from the source $x'$. The additional singularity in AdS space--time appears when $x$ sits on the light cone emanating from $\bar{x}'$ --- antipodal point of the source $x'$.} \cite{Avis1978AdS, Akhmedov2018Ultraviolet}. The extra singularity is due to the reflective boundary conditions. In fact, as we will see in the next section, the spectrum of modes is discrete and has an additional symmetry under the exchange $x \to \bar{x}$. 

As it is shown in \cite{Akhmedov2018Ultraviolet}, such a singularity may lead to the presence of non--local counterterms in an interacting theory if one adopts the standard Feynman's $i\ep$--prescription in the propagators for both the standard and additional UV singularities. Such an $i\ep$--prescription is due to an unusual time--ordering, which for the scalar field $\phi(x)$ and for the compact time coordinate in global AdS is as follows:

\begin{equation}
T \phi(x) \phi(x') = \theta\big[\sin(\tau-\tau')\big] \, \phi(x)\phi(x') + \theta\big[\sin(\tau'-\tau)\big]\, \phi(x')\phi(x).\label{tord}
\end{equation}
Such a time ordering respects the isometry in global AdS space--time if the points $x$ and $x'$ are timelike or lightlike separated. To show it one can rewrite the $\theta$-function in terms of embedding space coordinates: $\theta[\sin(\tau-\tau')] = \theta[X^{d-1} X'^{d} - X^{d} X'^{d-1}]$. This expression is invariant with respect to $SO(2,d-1)$ acting on $X^A$ if $|\zeta(x,x')| \le 1$. Due to $2\pi$-periodicity of $\sin(\tau-\tau')$ this expression is also invariant for timelike and lightlike separated points in CAdS. Such an invariance means that if $x \in C_n(x')$ (and therefore is timelike separated from $x$), then $\sign \sin(\tau-\tau') = (-1)^{n-1}$. This relation definitely holds if $x' = 0$~--- then $(n-1)\pi <|\tau| < n \pi$, and for generic $x'$ there is an isometry $g$ such that $0 = g x'$.

At the same time the usual time--ordering:

\begin{equation}
T_0 \phi(x) \phi(x') = \theta\big[\tau-\tau'\big] \, \phi(x)\phi(x') + \theta\big[\tau'-\tau\big] \, \phi(x')\phi(x)\label{tord2}
\end{equation}
is not isometry--invariant \cite{Castell1968Causality} for timelike separated points. In fact, the time coordinate in global AdS is compact: $\tau,\tau' \in [0,2\pi)$. A time translation $\tau \to (\tau + \Delta \tau) \text{ mod }2\pi$ can be represented as a rotation in the embedding Minkowski space--time and is, therefore, an isometry. It is easy to see that in general it does not preserve the ordering: if $\tau_1 + \tau_0 < 2\pi$ and $\tau_2 + \tau_0 > 2\pi$, then $\tau_2 > \tau_1$, but after the translation $(\tau_2 + \tau_0 - 2\pi) < \tau_1 + \tau_0$. Prescription (\ref{tord}) does not have this problem. Note that such a transformation also changes the relative region of $x$ and $x'$.

However, for any two points there exists a subset of isometries that preserves their relative region. It therefore also preserves the conventional time-ordering (\ref{tord2}) if $|\tau'-\tau| \ge |\theta'-\theta|$ (or equivalently $x' \notin S_0^+(x)$)~--- this isometry acts on $x$ and $x'$ the same way as in covering space. Hence, it does not change their relative region. The only region which has points with $\tau' > \tau$ and $\tau' < \tau$ (and so the order can be changed) is $S_0^+$. A general isometry can be represented as a composition of isometry from this subset and an additional $\tau$-translation  which moves either $\tau$ or $\tau'$ out of domain $[0,2\pi)$ and shifts $\tau - \tau' \to \tau - \tau' \pm 2\pi$.

\subsection{Free massive scalar field in 2D AdS space--time}\label{masst}

\noindent The Wightman function for the free scalar field theory in the global AdS space--time was calculated in \cite{Burgess1985Propagators} for $d > 2$ via the mode expansion. Because in 2D case the variable $\theta$ has different domain, it is instructive to do a separate calculation. We start from the following action:

\begin{equation}
S = \ili d^2x \sqrt{-g} \left[\dfrac{1}{2}(\p_\mu \phi)^2 - \dfrac{1}{2}m^2\phi^2\right] = \ili d\tau d\theta \left[\dfrac{1}{2}\eta^{\mu\nu} \p_\mu \phi \p_\nu \phi - \dfrac{1}{2} \dfrac{m^2}{\cos^2\theta}\phi^2\right],\label{massact}
\end{equation}
where we have used the explicit metric of the global AdS space--time (\ref{genmetr}) and $\eta_{\mu\nu}$ is the flat metric with Lorentzian signature $(+,-)$. 

This theory is equivalent to the one on the stripe with the $\theta$--dependent mass term. Hence, the mode decomposition of the field is as follows:

\begin{equation}
\phi(\tau,\theta) = \sum_{n=1}^{\infty} \bigg[a_n g_n(\tau,\theta) + \text{h.c.}\bigg]; \quad g_n(\tau,\theta) = e^{-i\w_n\tau} u_n(\theta),~n\in \mathbb{N},~\w_n>0,\label{phiop}
\end{equation}
where "h.c."\, stands for hermitian conjugate terms. The only nontrivial commutation relations for the creation and annihilation operators $a_n$ and $a_n'^{\dg}$ are $[a_n,a^\dg_{n'}] = \delta_{nn'}$. The functions $g_n$ should satisfy the equation of motion:

\begin{equation}
\<\square + \dfrac{m^2}{\cos^2\theta}\> g_n = 0,\label{KGg}
\end{equation}
where $\Box$ is the flat space--time d'Alembert operator.
Therefore, the equation for $u_n$ is as follows:
\begin{equation}
u_{n}'' + \w_n^2 u_{n} - \dfrac{m^2}{\cos^2 \theta} u_{n} = 0,\label{schrmass}
\end{equation}
where prime denotes the $\theta$--derivative. 

This equation can be interpreted as the stationary Schr{\"o}dinger equation for a particle of mass $1/2$ with energy $E_n = \w_n^2$ in the potential $V(\theta) = \frac{m^2}{\cos^2\theta}$. The potential is singular at $\theta = \pm \pi/2$. Due to the reflective boundary conditions described in section \ref{adsint}, namely $u_n(\pm \pi/2) = 0$, the spectrum is discrete.

We also have to impose the normalization conditions \cite{Birrell1982Quantum}:
\begin{equation}
(g_n,g_{n'}) = \delta_{nn'},\quad (g_n,g^*_{n'}) = 0,\quad (g^*_n,g^*_{n'}) = -\delta_{n n'},
\end{equation}
where we use the scalar product compatible with the equations of motion:

\begin{equation}
(\phi_1,\phi_2) = -i \ili_{\tau = \const} \Big[\phi_1(x) \p_\tau \phi_2^*(x) - \phi_2^*(x)\p_\tau \phi_1(x)\Big] \, d\theta.
\end{equation}
In terms of $u_n$ this means that
\begin{equation}
\braket{u_n}{u_{n'}} = \ili_{-\pi/2}^{\pi/2} u_n^*(\theta) u_{n'}(\theta) d\theta = \dfrac{\delta_{nn'}}{2\w_n}.\label{norm}
\end{equation}
Therefore, the modes are wave functions with the standard quantum mechanical normalization up to a constant factor.

It is convenient to reparametrize the mass in the standard way:
\begin{equation}
m^2 = s(s-1),~s>1.
\end{equation}
The substitutions $u_n = \cos^s\theta\,a_n$ and $y = \sin^2\theta$ bring the Schr{\"o}dinger equation into the standard hypergeometric form:

\begin{equation}
y(1-y)\p_y^2 a_n + \<\dfrac{1}{2} - (1+s)y\>\p_y a_n - \dfrac{1}{4}(s^2-\w_n^2) a_n = 0.
\end{equation}
Therefore, the general solution is as follows:

\begin{equation}
\begin{aligned}
u_{n} &=  C_1\cos^s \theta~_2F_1\<\frac{s-\w_n}{2},\frac{s+\w_n}{2};\frac{1}{2};\sin^2\theta\>+\\
&+ C_2 \sin\theta\cos^s\theta~_2F_1\<\dfrac{s-\w_n+1}{2},\dfrac{s+\w_n+1}{2};\dfrac{3}{2};\sin^2\theta\>,
\end{aligned}
\end{equation}
where$~_2F_1$ is the hypergeometric function. 

This solution should be regular at $\theta = \pm \pi/2$, i.e. at $\sin^2\theta = 1$. The hypergeometric functions are singular at this point except for the case when one of their first two arguments is negative integer. Otherwise one can show using their transformation relations that $u_n \sim \cos^{1-s}\theta$, as $\theta \to \pm \pi/2$, which is singular (see e.g. \cite{Akhmedova:2019bau}). 

As $\w_n> 0$ and $s > 1$ we find that either $\w_n-s = 2k$ or $\w_n-s = 2k+1$, $k \in \mathbb{N}\cup \{0\}$. Using the relation between the hypergeometric functions and Jacobi polynomials $P_n^{(\alpha,\beta)}$, one can rewrite the modes as:

\begin{eqnarray}
\w_n = s+n-1, \quad u_{2n-1} = A_n \cos^s\theta P_{n-1}^{(-1/2,\,s-1/2)}(\cos 2\theta), \nonumber \\ u_{2n} = B_{n} \cos^s\theta \sin\theta P_{n-1}^{(1/2,\,s-1/2)}(\cos 2\theta).\label{massspectr}
\end{eqnarray}
The normalization constants $A_n$ and $B_n$ and the Wightman function are calculated in appendix B. These modes are quasiperiodic:
\begin{equation}
g_n(\bar{x}) = e^{-i\pi s} g_n(x).\label{modtrans}
\end{equation}
They are single-valued on the CAdS only when $2\pi$-periodic in $\tau$. This condition is satisfied if $s \in \mathbb{N}$~--- the modes can be defined in global AdS space only for the discrete spectrum of masses. We will, however, consider general values of $s$ to obtain the Wightman function on the CAdS, where there are no such restrictions.

With the modes under consideration one can calculate the Wightman function $W(x,x')$ over the Fock vacuum, where $x = (\tau,\theta)$, $x' = (\tau',\theta')$. The vacuum is defined as a state which is annihilated by all $a_n$ operators. Then the expression for the given mass, i.e. $s$, is as follows:

\begin{equation}
W^{(s)}(x,x') = \bl \phi(x) \phi(x') \br = \sum_{n=1}^{\infty} g_n(x) g^*_n(x') = \sum_{n=1}^{\infty} e^{-i\w_n(\tau-\tau' - i\ep)}u_{n}(\theta)u^*_{n}(\theta'),\label{wightsum}
\end{equation}
here $i\ep$-term is added for the convergence in singular points. This function is quasiperiodic due to the quasiperiodicity of modes~--- this fact leads to additional singularities. In appendix B we find that:

\begin{equation}
W^{(s)}(x,x') = \dfrac{1}{2\pi} Q_{s-1}\big[\zeta + i\ep \sin(\tau-\tau')\big], \label{masswig}
\end{equation}
where $Q_{s-1}$ is the Legendre function of the second kind and $\zeta$ is the invariant defined in (\ref{gpar}). When $|\tau - \tau'| < |\theta-\theta'|$ (in this region $\zeta > 1$) the value is the same as of the standard Legendre function defined on the plane with cut at $(-\infty,1)$. This result coincides with the one obtained in \cite{Akhmedov2018Ultraviolet} via a different calculation. 

After the translation $\tau \to \tau + 2\pi$ the argument of the Legendre function $(\zeta + i \epsilon \sin (\tau - \tau'))$ encircles the line segment $[-1,1]$ counterclockwise. Due to monodromy properties of $Q_{s-1}$, it gets multiplied by $e^{-2i\pi s}$. That agrees with the relation (\ref{modtrans}). Hence, the RHS of equation (\ref{masswig}) is well-defined on Riemann surface, which is fibering over the complex $\zeta$--plane with the cut along the $[-1,1]$ line, such that each time translation by $2\pi$ lifts up to a new sheet. Such a behavior and the fact that $\tau-\tau'$ on the global AdS varies from $-2\pi$ to $2\pi$ (and hence the segment $[-1,1]$ is encircled two times) means that the Wightman function is not invariant with respect to global AdS isometries, unless $s \in \mathbb{N}$.

Let us show that this function is, however, invariant with respect to the isometries of the CAdS space--time. If there is the isometry violating the invariance, it should move $\zeta$ to another sheet of the Riemann surface. But the sheet is uniquely determined by the relative region of $x$ and $x'$~--- one can see it directly from eq. (\ref{masswig}) and the constancy of $\sign \sin (\tau-\tau')$ when $x\in C_n(x')$, i.e. when $|\zeta|<1$.  As the relative region does not change under isometries, the sheet cannot change as well and, hence, the Wightman function is invariant. This invariance  is a rather expected outcome, as isometries of the CAdS space--time are symmetries of the action.

Using ordinary time ordering prescription (\ref{tord2}), we can find the expression for the Feynman propagator $\tilde{F}^{s}(x,x')$ on the CAdS space--time:

\begin{equation}
\tilde{F}^{(s)}(x,x') = \begin{cases}
\frac{e^{-2i\pi n s}}{2\pi} Q_{s-1}(\zeta + i\ep),\quad n < \frac{|\tau-\tau'|}{2\pi} < n+\frac{1}{2};\\\\
\frac{e^{-2i\pi (n+1) s}}{2\pi} Q_{s-1}(\zeta - i\ep),\quad n + \frac{1}{2} < \frac{|\tau-\tau'|}{2\pi} < n+1,
\end{cases}\label{massprop}
\end{equation}
where $n \in \mathbb{N} \cup \{0\}$. In this formula $Q_{s-1}$ is defined on the complex $\zeta$--plane with the standard for Legendre functions cut at $(-\infty,1)$. This function is also isometry-invariant on the CAdS space--time: when $x' \notin  S_0^+(x)$ the standard time-ordering is invariant, hence, the whole function is invariant due to the invariance of the Wightman function. And when $x' \in S_0^+(x)$ we have $|\tau-\tau'| < \pi$, so the Feynman propagator is given by $\frac{1}{2\pi} Q_{s-1}(\zeta+i\lc)$, which is clearly invariant.

As the Wightman function is global AdS isometry-invariant only when $s \in \mathbb{N}\cup \{0\}$ (in other cases it is not periodic), the causal propagator constructed with AdS time-ordering is also invariant only for these values of $s$. However, it is still possible to formally define such a propagator in global AdS. Namely, we define it as:
\begin{equation}
F^{(s)}(x,x') = \dfrac{1}{2\pi}Q_{s-1}(\zeta + i\ep).\label{massfeyn}
\end{equation}
As it is discussed in section \ref{cvpnk}, this propagator can be associated with the theory of the same mass defined in Poincare patch.

In section \ref{pert} we use the propagator for massless theory $m=0$, i.e. $s = 1$. Hence, let us consider this case separately. The frequencies, normalized modes and the Wightman function are as follows:

\begin{equation}
\w_n = n,\quad u_n = \dfrac{1}{\sqrt{\pi n}}\sin\left[n\<\theta-\dfrac{\pi}{2}\> \right],\quad W(x,x') = \dfrac{1}{4\pi}\log\<\dfrac{\zeta+1 + i\ep \sin (\tau-\tau')}{\zeta-1 + i\ep\sin(\tau-\tau')}\>.\label{wig}
\end{equation}
In this case we can define the Feynman propagator using the invariant time-ordering (\ref{tord}):
\begin{equation}
F(x,x') = \dfrac{1}{4\pi}\log\<\dfrac{\zeta+1 + i\ep }{\zeta-1 + i\ep}\>.\label{fprop}
\end{equation}
This expression coincides with the formally defined periodic propagator (\ref{massfeyn}), when $s=1$.

\subsection{Massive field in a general static asymptotically AdS manifold}\label{massgen}

\noindent In this section we consider the static asymptotically AdS manifold (\ref{genmetr}). We restrict our attention to the case when the conformal factor $f(\theta)$ behaves as in eq. (\ref{f}). Our goal is to find out how generic is the phenomenon of the presence of the non--local UV singularities in the propagators. We will show here that there is the UV singularity localized on the light--cone emanating from the antipodal point when $\alpha \leq 2$. 

Now instead of (\ref{massact}) the action is:

\begin{equation}
S = \ili d\tau d\theta \left[\dfrac{1}{2} \eta^{\mu\nu} \p_\mu \phi \p_\nu \phi - \dfrac{1}{2} m^2 f(\theta)\phi^2\right],
\end{equation}
where the function $f$ satisfies the condition (\ref{f}). Again $m$ can be absorbed into $f$ so we can set $m = 1$. 

The mode decomposition and the normalization conditions, (\ref{phiop}) and (\ref{norm}), remain the same. The equations of motion for the modes are as follows:

\begin{equation}
u_{n}'' + \w_n^2 u_{n} - f(\theta)u_{n} = 0.
\end{equation}
Now the ``potential'' $f(\theta)$ in this equation grows to infinity at the boundary of the region $|\theta| < \pi/2$. Hence, naively the spectrum is discrete. However, we will see in a moment that the situation is a bit more tricky for the case when $\alpha \leq 2$.

This equation cannot be solved exactly for general $f$, but the main contribution to the UV singularities of the Wightman function comes from high--frequency modes. Hence, below we look for the high--energy asymptotic behaviour of the modes.

As we have observed in subsection \ref{GFoc}, at zeroth order the effect of generic $f$ can be approximated by the infinite potential well. Then the situation is analogous to the case of massless field \ref{wig}, where

\begin{equation}
\w_n = n + \Delta_n;~|\Delta_n| \ll n,~n\to\infty.\label{dndef}
\end{equation}
However, it is still possible that $|\Delta_n|$ is non-vanishing as $n\to \infty$. For example, from (\ref{massspectr}) one can find that for the massive field in AdS $|\Delta_n| = s-1$. Such a contribution to $\omega_n$ does affect the singular behaviour, as we will see below. Therefore we should take it into account. Our goal now is to find it for generic $f(\theta)$. 

More accurate approximation for the frequencies can be found from the semiclassical (WKB) method. The WKB action in this case coincides with the action (\ref{contS}). Therefore, the quantization rule for WKB frequencies $\tilde{\w}_n$ is as follows:

\begin{equation}
\oint \sqrt{\tilde{\w}_n^2-f(\theta)}d\theta = 2\pi (n + \gamma),\label{tildeWKB}
\end{equation}
where $\gamma$ is the Maslov index \cite{Maslov1965Asymptotic}. For very large values of $n$ the difference between the exact $\omega_n$ and semiclassical $\tilde{\omega}$ is vanishing.
Hence, the WKB spectrum is sufficient to determine the most singular part of the correlation function. 

Thus, to continue we can replace $\tilde{\w}_n$ with $\w_n$ in (\ref{tildeWKB}):

\begin{equation}
S_n = \oint \sqrt{\w_n^2-f(\theta)}d\theta = 2\pi (n + \gamma).\label{WKB}
\end{equation} 
It is probably worth stressing here that when $\alpha < 2$ (see (\ref{f})) the modes, which grow in the classically forbidden zone, also become normalizable. This means that the potential barrier in the region $[\theta_t,\pi/2]$ becomes penetrable as the integral for the action on this interval is convergent. As a result, the spectrum becomes continuous. Therefore, to restrict ourselves to the asymptotically AdS case we need to impose the Dirichlet boundary conditions $u_n(\pm \pi/2) = 0$. That makes the spectrum discrete for any $\alpha$. 

Furthermore, the WKB approximation is applicable when $|\lambda'(\theta)| \ll 1$, where $\lambda = \frac{1}{\sqrt{\w_n^2 - f(\theta)}}$ is the de Broglie wavelength. Near the boundaries $\pm \pi/2$ this condition reduces to

\begin{equation}
(\pi/2 \mp \theta)^{\frac{\alpha}{2} - 1} \ll 1,\quad \theta \to \pm \pi/2.
\end{equation}
It is satisfied if $\alpha > 2$, so in this case WKB wave functions can be used in the forbidden region $|\theta| > \theta_t$. In this case the characteristic scale of the significant change of the potential near the turning point,

\begin{equation}
\Delta \theta = \left| \dfrac{f'(\theta_t)}{f(\theta_t)} \right| \sim \dfrac{\pi}{2} - \theta_t, \label{dthet}
\end{equation}
is much greater than the scale $a$, where WKB becomes valid ($|\lambda'(\theta_t+a)| \sim 1$),

\begin{equation}
a =  \<\dfrac{\pi}{2} - \theta_t\>^{\frac{\alpha+1}{3}}.
\end{equation}
Here we have used the linear approximation of the potential. As there is a region where both WKB and linear approximation work, the Maslov index is the standard one: $\gamma = -\frac{1}{2}$. It differs from usually used $1/2$ (e.g. \cite{Landau1981Quantum}) as we count states starting from $n = 1$ rather than $n = 0$.


In the case $\alpha < 2$ the WKB method cannot be used in the classically forbidden zone near the boundaries. However, in this case near the turning points $\theta_t$ the de Broglie wavelength $\lambda \sim (\pi/2-\theta_t)^{\frac{\alpha}{2}}$ is much greater than $\pi/2-\theta_t$. Then the wave function does not change significantly on the scale $\Delta \theta$ and for high energy modes we can ignore the presence of the potential. Hence, the infinite potential well approximation (for which $\gamma = 0$) can be used to define the finite part of $\w_n$. The most singular part of the Wightman function is then exactly the same as in the massless theory.

The case $\alpha = 2$ is the most non-trivial. The WKB approximation can be used only far from the turning point, when $f(\theta)$ can be neglected. Near the boundary we have to solve the Schr{\"o}dinger equation with the potential given by (\ref{f}). For large values of $\omega_n$ the solution can be glued to the one of the WKB approximation. Hence, $\gamma$ depends only on $C$ from eq. (\ref{f}). It means that we can use the exact result for massive theory in exact AdS to calculate $\gamma$, as in this case $\alpha = 2$.
Then the spectrum is given by (\ref{massspectr}), and the action by (\ref{foccond2}) with $\delta = -2\pi m = -2\pi \sqrt{C}$.

In all, we obtain the following possible values of $\gamma$ in (\ref{WKB}) depending on the value of $\alpha$:

\begin{equation}
\gamma = \begin{cases}
-\frac{1}{2},&\alpha > 2;\\
0,& \alpha < 2;\\
s-1-\sqrt{C},& \alpha = 2,
\end{cases}\label{Maslov}
\end{equation}
where $s = \frac{1+\sqrt{1+4C}}{2}$, because $s(s-1) = C$.

Before continuing with the full solution of the WKB approximation (\ref{WKB}), let us establish some connection with the considerations of section \ref{GFoc}. Let us assume that $\w_n = n + \beta$, $\beta > 0$. It means that the WKB modes are quasiperiodic in the sense of (\ref{modtrans}) with $s = \beta+1$. The equation (\ref{WKB}) turns into the geodesic focusing condition (\ref{foccond2}) with $\delta = 2\pi(\gamma-\beta)$ for $\w = \w_n$. Then, if (\ref{foccond2}) is satisfied for a set $\{\w_n\}$, $n \in \mathbb{N}$, it is also satisfied for each $\w > 0$ because $S(\w)$ is a monotonically increasing function. Therefore, the existence of geodesic focal point is equivalent to quasiperiodicity of WKB modes. However, we have observed that in the case of exact global AdS the exact modes are also quasiperiodic, but with different phase multiplier. We have also shown that global AdS space--time is the only manifold with symmetric $f(\theta)$ having the geodesic focusing property by solving the inverse problem~--- deduction of $f(\theta)$ from $S(\w)$. The analogous problem in quantum case is to deduce the potential from the exact spectrum $\w_n$. It is much more complicated, but for symmetric potentials it is possible to prove (see e.g. \cite{Freiling2001InverseSP}) that the solution is unique. Therefore, the only asymptotically AdS manifold with the spectrum $\w_n = n+\beta, \quad \beta > 0$ is AdS space itself, because according to (\ref{massspectr}) all such spectra were obtained from massive field theories in AdS space. 

Thus, for the case $\alpha < 2$ we have seen that the infinite well approximation can be used and there is the non--local UV singularity, as in massless case. Now we need to estimate the LHS of (\ref{WKB}) when $\alpha \ge 2$. Below in this section we will omit all quantities vanishing in the $n\to\infty$ limit. Let us use the series expansion:
\begin{equation}
S_n \approx \w_n\oint \<1 - \sum_{k=1}^{\infty} \dfrac{(2k-3)!!}{k! 2^k} \<\frac{f(\theta)}{\w_n^2}\>^{k}\>d\theta.
\end{equation}
The first term here is $4\theta_t \w_n$. Using the expression (\ref{turn}) for turning points, we find that:
\begin{equation}
4\theta_t \w_n \approx 2\pi \w_n - 4 C^{\frac{1}{\alpha}} \w_n^{1-\frac{2}{\alpha}}.\label{dsfrst}
\end{equation}
The second term can give non-vanishing contribution only near turning points where $f(\theta) \sim \w_n^2$. The characteristic size of such area is $\Delta\theta \sim 1/\w^{\frac{2}{\alpha}}$ defined in (\ref{dthet}), the terms of the sum have zero order in $\w_n$ near $\theta_t$. Hence, their contribution is of the same order as $-\w_n \Delta \theta \sim -\w_n^{1-\frac{2}{\alpha}}$. Combining with (\ref{dsfrst}), we find:
\begin{equation}
S_n \approx 2\pi \w_n - D \w_n^{1-\frac{2}{\alpha}},\quad D > 0.\label{D}
\end{equation}
The correction is non-vanishing when $\w_n \to \infty$ if $\alpha \ge 2$. In this case it is possible to show that $D$ depends only on an asymptotic behavior of its potential and is equal to:
\begin{equation}
D = 2 \sqrt{\pi} C^{\frac{1}{\alpha}}\dfrac{\Gamma\<\frac{\alpha-1}{\alpha}\>}{\Gamma\<\frac{3\alpha-2}{2\alpha}\>}.\label{coefD}
\end{equation}
This value is found in appendix \ref{WKBcor}. 

As a result, the WKB mode frequencies are as follows:
\begin{equation}
\w_n \approx n+ \gamma + \dfrac{D}{2\pi} n^{1-\frac{2}{\alpha}}.\label{WKBspec}
\end{equation}
Because the potential $f(\theta)$ is an even function the modes are either even or odd functions. According to the oscillation theorem \cite{Landau1981Quantum}, the wave function of $n$-th state has $n-1$ zeros. Hence, it is symmetric for even $n$ and antisymmetric for odd. Near the origin the potential can be neglected, hence for $n \gg 1$ we can use the following approximation
\begin{equation}
u_n \approx \dfrac{1}{\sqrt{\pi n}} \sin\<\w_n\theta -\dfrac{\pi n}{2}\>.
\end{equation}
We use the zeroth-order approximation for the normalization coefficients, i.e. take them the same as in (\ref{wig}). Corrections to the latter after substitution into the expression for Wightman function (\ref{wightsum}) do not create singular terms~--- the corresponding series are absolutely convergent. Therefore, near the singularities the Wigtman function $W_f(x,x')$ can be expressed as follows:
\begin{equation}
\begin{aligned}
W_f(x,x') &\approx \sum_{n} e^{-i\w_n(\tau-\tau' - i\lc)} \sin\<\w_n\theta -\dfrac{\pi n}{2}\>\sin\<\w_n\theta' -\dfrac{\pi n}{2}\> = \\
&=\sum_n \dfrac{1}{4\pi n}\<e^{-i\w_n(\tau-\tau'-i\lc - (\theta -\theta'))} + e^{-i\w_n(\tau-\tau'-i\lc + (\theta -\theta'))} + \right.\\
&+\left. (-1)^ne^{-i\w_n(\tau-\tau'-i\lc - (\theta +\theta'))} + (-1)^n e^{-i\w_n(\tau-\tau'-i\lc + (\theta +\theta'))}\>,
\end{aligned}
\end{equation}
$\w_n$ here are WKB frequencies. It is always singular when $\tau - \tau' = \pm(\theta-\theta')$~--- the usual local singularity. Other singularities can be located at $\tau - \tau' = 2\pi l \pm(\theta-\theta')$ and $\tau - \tau' = 2\pi(l-1/2) \pm (\theta + \theta')$ for $l \in \mathbb{Z}$ as $\w_n = n+\Delta_n \sim n$ for large $n$. However these points are singular if the series
\begin{equation}
\mt{S} = \sum_n e^{i \pi l \Delta_n - \lc n},\quad l\neq 0.
\end{equation}
is divergent after taking the limit $\lc \to 0$. According to the formula for the frequencies (\ref{WKBspec}),
\begin{equation}
\mt{S} \approx \sum_n \exp\left[ i \pi l\<\gamma + \frac{D}{2\pi} n^{1-\frac{2}{\alpha}}\> - \lc n \right].
\end{equation}
It is divergent only for $\alpha \le 2$ and the additional singularities in this case have the phase multipliers. Using the expressions (\ref{coefD}) and (\ref{Maslov}) for $D$ and $\gamma$ respectively one can show that they coincide with the phase multiplier in massive theory in AdS with $m = \sqrt{C}$. 

To summarize: for $\alpha < 2$ the singularities of the Wightman function are the same as in massless theory in AdS, for $\alpha = 2$~--- as in massive theory in AdS with mass $m = \sqrt{C}$ and for $\alpha > 2$ there is only the standard local UV singularity.

\subsection{Higher dimensions}

\noindent To conclude this section we discuss the generalization of the above observations to the dimensions higher than two. We consider a $d$-dimensional asymptotically AdS space--time

\begin{equation}
ds_d^2 =  f(\theta)\<d\tau^2-d\theta^2 - \sin^2\theta\,d\Omega_{d-2}^2\>,
\end{equation}
where $f(\theta)$ satisfies (\ref{f}). The Weyl transformation now acts on the field non-trivially: the conformal weight of a scalar field is $\frac{d-2}{2}$. Then the transformation $\phi \to f^{\frac{2-d}{4}} \phi$ changes the action into the following form:

\begin{equation}
S = \ili d^d x \sqrt{-h} \left[ \dfrac{1}{2} h^{\mu\nu} \p_\mu \phi \p_\nu \phi - \dfrac{d-2}{8} \phi^2\<\dfrac{f''}{f} + \dfrac{d-6}{4} \<\dfrac{f'}{f}\>^2\> - \dfrac{1}{2}f m^2 \phi^2 \right],
\end{equation}
where $h$ is a metric on ESU. The second term here has quadratic singularity at $\theta = \pi/2$, hence it can be combined with the third one into $\frac{1}{2}g \phi^2$. 
The equation of motion for the modes (\ref{phiop}) where $u_n = u_\w(\theta,\Omega_{d-2})$ is as follows:
\begin{equation}
\p^2_\theta u_n + (d-2)\cot \theta\, \p_\theta u_n + \dfrac{1}{\sin^2\theta}\Delta_{d-2} u_n + \w^2 u_n- g(\theta) u_n = 0,
\end{equation}
where $\Delta_{d-2}$ is a Laplace operator on $S^{d-2}$. The first $\theta$-derivative can be eliminated via substitution $u_n = v_n(\sin\theta)^{\frac{2-d}{2}}$:
\begin{equation}
\p^2_\theta v_n +\dfrac{1}{\sin^2\theta}\Delta_{d-2} v_n + \<\w^2 - g(\theta) + \dfrac{d-2}{2} - \dfrac{(d-2)(d+4)}{4}\cot^2\theta\>v_n = 0;
\end{equation}
$u_n$ can be chosen as an eigenfunction of $\Delta_{d-2}$. As a result, we obtain one-dimensional Schr{\"o}dinger equation. The only term singular at $\theta = \pi/2$ is $g(\theta)$, at $\theta = 0$ the singularities are quadratic. Hence, this problem can be treated similarly to the two-dimensional case. 

The calculation of the Maslov index in this case is much more complicated, but the divergent term in $\Delta_n$ which is defined similarly to (\ref{dndef}) (in the conformally coupled theory the frequencies are integer \cite{Burgess1985Propagators}) can be estimated. As the difference between $f$ and $g$ has quadratic singularity at $\pi/2$, calculations similar to $d=2$ case show that such divergent term does exist also when $\alpha > 2$~--- the same condition as before. However the local singularity should be $\sim \frac{1}{\<s(x,x')\>^{d-2}}$ rather than $\log s(x,x')$ as in two-dimensional case, where $s(x,x')$ is a geodesic distance between $x$ and $x'$. It means that non--local UV singularity does appear for every potential, but for $\alpha > 2$ it is of lower order than the local one.

This argument can be extended to general manifolds with boundary where the metric is singular. Namely, the structure of singularities depends only on the behavior of high-frequency modes near the boundary. Hence we can conclude that the non-local singularities are absent if and only if the singularity of metric at the boundary is stronger than quadratic.

\section{Principal chiral field and perturbation theory in global AdS}\label{pert}

\noindent So far we have considered only free (Gaussian) theories over asymptotically  AdS space--times. To see the consequences of the second UV singularity in the propagators in this section we consider an interacting theory in global AdS. We will find that the second singularity leads to the appearance of  non-local counterterms \cite{Akhmedov2018Ultraviolet}. We will make a few comments on other difficulties in global AdS and then calculate the effective action for principal chiral field theory.

\subsection{Problems with perturbation expansion in global AdS}

\noindent Consider an interacting scalar field theory in global Lorentzian AdS space--time. Besides the boundary, global AdS also has closed timelike geodesics. As we already pointed out in section \ref{adsint}, this property means that the time ordering cannot be defined in usual way. The isometry invariant definition is given by (\ref{tord}). 
The problem with this definition is that it is not transitive, i.e. one cannot use it for more than two fields. 

Instead of that, for the free theory one can use Wick's theorem as a definition. However, it does not work for interacting fields. We can only define the perturbative expressions using Feynman rules, but this approach is in some way inconsistent. To show the latter point, let us consider the massless free field theory:

\begin{equation}
S = \ili d^2x \sqrt{-g} \dfrac{1+\lc}{2} (\p_\mu \phi)^2,\quad \lc > 0.
\end{equation}
By the field redefinition $\phi \to \frac{\phi}{\sqrt{1+\lc}}$ this action obviously can be transformed into the standard form (\ref{massact}). Hence, the Feynman propagator $\mt{F}(x,x';\lc)$ should be as follows:

\begin{equation}
\mt{F}(x,x';\lc) = \dfrac{F(x,x')}{1+\lc},\label{epspr}
\end{equation}
where $F(x,x')$ is defined in (\ref{fprop}). 

On the other hand, one should be able to calculate the propagator treating $\frac{\lc}{2}(\p_\mu \phi)^2$ as a perturbation. The first-order correction in $\lc$ is as follows:

\begin{equation}
\mt{F}^{(1)}(x,x') = i\lc \ili d^2y \sqrt{-g(y)} \p_\mu^y F(x,y) \p^{y\mu} F(y,x'),\label{1corr}
\end{equation}
here $\p^y_\mu = \frac{\p}{\p y^\mu}$. Integrating by parts, we get:

\begin{equation}
\mt{F}^{(1)}(x,x') = -i\lc \ili d^2y \sqrt{-g(y)} F(x,y)\Box_y F(y,x') + i\lc \ili_{\p \text{AdS}} F(x,y) \p_\mu F(y,x') dy^\mu,
\end{equation}
where $\Box_y = \frac{1}{\sqrt{-g}} \p_\mu (\sqrt{-g} \p^\mu)$ and $\p\text{AdS}$ is the boundary of AdS at $\theta = \pm\pi/2$. Using (\ref{fprop}), one can show that the boundary term is zero. The function $F(y,x')$ satisfies the wave equation almost everywhere except for $y^0-x'^0 = 0$ and $|y^0 - x'^0| = \pi$, where the step-functions in the definition (\ref{tord}) have discontinuities. From the canonical commutation relations and the quasiperiodicity (\ref{modtrans}) for massless theory we obtain:

\begin{equation}
\Box_y F(y,x') = -i\delta_{\text{AdS}}(y,x') - i\delta_{\text{AdS}}(y,\bar{x}'),\label{KGeq} 
\end{equation}
here $\delta_{\text{AdS}}(x,y)$ is the delta-function for the measure $\sqrt{-g}\,d^2y$ and $\bar{x}'$ is the antipodal to $x$ point. Hence, it follows that

\begin{equation}
\mt{F}^{(1)}(x,x') = -\lc \Big[F(x,x') + F(x,\bar{x}')\Big].
\end{equation}
The RHS of this expression can be rewritten in terms of the commutator $C(x,x') = [\phi(x),\phi(x')]$:

\begin{equation}
\mt{F}^{(1)}(x,x') = -\lc\Big\{\theta[\sin(\tau-\tau')]C(x,x') +\theta[\sin(\tau'-\tau)]C(x',x)\Big\},
\end{equation}
where we have used that $W(x,x') = -W(\bar{x},x')$. This result does not coincide with the first-order expansion following from (\ref{epspr}). 

The correct answer can be obtained by replacing $F(y,x')$ with the function $F_P(y,x')$ which satisfies (\ref{KGeq}) with only local delta-function on the RHS, which is zero at the boundary and is $2\pi$-periodic in $y^0$. The latter condition is required to avoid the boundary terms at $y^0 = \pm \pi/2$. Such a propagator is constructed in the next subsection. However, it turns out to be ill-defined in the massless case and for all massive theories with isometry-invariant Wightman functions. There is also a possibility to consider the theory in the covering manifold with $\tau \in (-\infty,+\infty)$ and restore the $i\ep$ prescription after calculations, but the propagator does not decay as $\tau \to \pm \infty$.  We will come back to the discussion of this point in the next section.


\subsection{Another way to define the propagator in global AdS}

\noindent Yet another way to define the propagator in global AdS is with the use of the path integral:

\begin{equation}
F_P(x,x') = \dfrac{\ili \mt{D} \phi\, \phi(x)\phi(x') e^{iS[\phi]}}{\ili \mt{D} \phi\,e^{iS[\phi]}},
\end{equation}
with the periodic time $\tau \in [0,2\pi)$ and periodic boundary conditions for the field. Let us be more specific. A natural way is to impose $\phi(\tau = 0) = \phi(\tau = 2\pi)$, the reflective boundary conditions are not essential here. In the Hamiltonian formalism the expression under consideration is as follows:

\begin{equation}
F_P(x,x') = \dfrac{\Tr\<T_0 \{\phi(x)\phi(x') \}e^{- 2i\pi H}\>}{\Tr e^{-2i\pi H}},\label{temp}
\end{equation}
where $T_0$ denotes the usual time-ordering, as we have already denoted it above. 

As $\tau$-interval is finite, the contribution of excited states is not suppressed. Hence, the expression under consideration obviously does not define the average over the vacuum state. It resembles the Matsubara Green function but with imaginary temperature $2i\pi$. The main advantage of this expression is that it is manifestly $2\pi$-periodic in Lorentzian time $\tau$. Moreover, (\ref{temp}) is isometry-invariant~--- the measure is considered to be invariant and the periodic boundary conditions are consistent with the invariance. Hence, this propagator is well-defined in global AdS, unlike the one over the vacuum state. 

Furthermore, in the free theory $F_P(x,x')$ does satisfy eq. (\ref{KGeq})
with only one local delta-function on the RHS. That is due to the conventional time-ordering. Using the mode decomposition (\ref{phiop}) and the standard basis of Fock states, one can obtain the explicit expression:

\begin{equation}
F_P(x,x') = \tilde{F}(x,x') + \sum_\w \dfrac{1}{e^{2\pi i \w} - 1}\Big[h_\w(x) h^*_\w(x') + h^*_\w(x)h_\w(x')\Big],
\end{equation}
here $\tilde{F}(x,x')$ is a vacuum expectation value of $T_0 \{\phi(x)\phi(x') \}$. Note that it is ill-defined when at least one of $\omega$ is integer.

In the case of two-dimensional massive theory with $m^2 = s(s-1)$, $s>1$ the spectrum is defined by (\ref{massspectr}): $\w_n = s+n-1$, $n\in\mathbb{N}$. Such a simple form allows for the exact calculation of this ``thermal'' propagator:
\begin{equation}
F_P^{(s)} = \tilde{F}^{(s)}(x,x') + \dfrac{W^{(s)}(x,x') + W^{(s)}(x',x)}{e^{2\pi i s}-1},\label{masstherm}
\end{equation}
where $\tilde{F}^{(s)}$ is defined in (\ref{massprop}) and $W^{(s)}$ in (\ref{masswig}). Using the definition
\begin{equation}
\tilde{F}^{(s)}(x,x') = \theta(\tau-\tau')W^{(s)}(x,x') + \theta(\tau'-\tau)W^{(s)}(x',x),
\end{equation}
and the monodromy properties established in subsection \ref{masst}  one can show that $F_P^{(s)}(x,x')|_{x^0 = 0} = F_P^{(s)}(x,x')|_{x^0 = 2\pi}$, i.e. it is periodic. 

To prove the isometry invariance of the propagator in question we refer to the discussion in the beginning of section \ref{nlsing}. We have already seen that the propagator $\tilde{F}^{(s)}$ and the Wightman functions are invariant under the action of a subset of isometries preserving the relative region of $x$ and $x'$ (as they belong to the isometry group of the CAdS space--time). Because any other isometry can be obtained by additional time translation shifting $\tau-\tau' \to \tau-\tau' \pm 2\pi$, the invariance with respect to the whole group follows from $2\pi$-periodicity of the function $F_P^{(s)}$~--- the Wightman function is not invariant separately due to its non-trivial monodromy. Another feature of this propagator is that it cannot be defined as a boundary value of some analytic function of $\zeta$~--- the functions $W^{(s)}(x,x')$ and $W^{(s)}(x',x)$ are defined on different parts of Riemann surface.

\subsection{Effective action for principal chiral field}\label{EfPCF}

\noindent To show the consequences of the non--local UV singularity and of different ways of defining propagators, we consider the principal chiral field theory in 2D global AdS space--time. Our goal is to calculate the one loop UV counter--terms. 

The tree--level action is as follows:

\begin{equation}
S = - \dfrac{1}{2e_0^2} \int d^2x\,\sqrt{g}\, \Tr R_\mu^2,\quad R_\mu = \grg\, \p_\mu \grg^{-1},
\end{equation}
where $\grg(x) \in G$~--- a compact simple Lie group. The standard method is to decompose the field into the classical and quantum components and then integrate out the latter. Substituting $\grg(x) = h(x) \grg^{cl}(x)$, where $\grg^{cl}$ is a classical field and $h = e^\phi,~\phi \in \text{Lie}(G)$ is the quantum one. Expanding the action up to the second--order terms in $\phi$, we get :

\begin{equation}
S = \dfrac{1}{2e_0^2} \int d^2x\,\sqrt{g}\, \Tr \left(-R_\mu^{cl\,2} +\phi\,\Box \,\phi -R_\mu^{cl} [\phi, \p^\mu \phi]\right),
\end{equation}
where $\Box$ is the Laplace-Beltrami operator in AdS space--time. We also omitted the first-order terms as they generate only tadpole diagrams. 

Let us choose a basis $t^I$ in the Lie algebra ($\phi = i\phi^I t^I$), for which the Killing form is diagonal, and

\begin{equation}
\Tr t^I t^J = \dfrac{1}{2} \delta^{IJ},\quad \,[t^I, t^J] = i f^{IJK} t^K.
\end{equation}
After the additional rescaling $\phi^I \to \sqrt{2e_0^2}\phi^I$ we obtain:
\begin{equation}
S = S_{\text{cl}} + \dfrac{1}{2} \int d^2x\, \sqrt{g}\,\phi^I \left( \delta^{IJ} \Box - i f^{IJK}  R_\mu^{cl\,K} \p^\mu \right) \phi^J,
\end{equation}
where $S_{\text{cl}} = S[\grg^{\text{cl}}]$.
 
The free theory of quantum field $\phi^I$ is massless. Hence, the periodic propagator (\ref{temp}) is infinite and thus cannot be defined. It is possible to use the embedding space path integral with $\tau \in (-\infty,\infty)$, but in the path integral approach Feynman propagator has conventional time-ordering and therefore it is not isometry-invariant, as we have discussed above. 

\begin{figure}
\centering\includegraphics[width=0.85\textwidth]{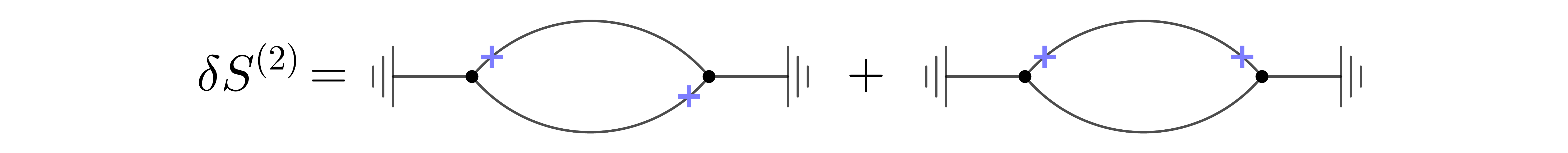}
\caption{The second-order correction to the effective action} \label{Diag}
\end{figure}

The isometry-invariant propagator $F^{IJ}(x,x')$ can be defined via AdS time-ordering. Due to the simple structure of colour indices of the free field term, this propagator is given by $F^{IJ}(x,x') = \delta^{IJ} F(x,x')$, where $F(x,x')$ is given by (\ref{fprop}):
\begin{equation}
F(x,x') = \dfrac{1}{4\pi} \left[\log \dfrac{1}{\zeta - 1 + i \epsilon} - \log \dfrac{1}{\zeta + 1 + i \epsilon} \right].
\end{equation}
Because both singularities have the Feynman $\epsilon$--prescription, we expect the appearance of the non-local counter--term \cite{Akhmedov2018Ultraviolet}. Fig. \ref{Diag} shows the diagrams which provide the first loop correction. The external lines stand for the classical field $R_\mu^{\text{cl}K}$ and the crosses on the internal lines denote the derivatives $\p_\mu$. 

Via integration by parts one can shows that the contributions of the two diagrams coincide if $\nabla_\mu R^{\text{cl}\,\mu} = 0$, which is the classical equation of motion. However, in general case the difference between them is UV-finite, because the integrand has only one of the derivatives in the integral acting on the propagator.
In fact, in general case, the difference between contributions of two diagrams is proportional to 

\begin{equation}
    \ili d^2x\,d^2x' \sqrt{-g(x)} \sqrt{-g(x')} R_\mu^{\text{cl}\,I}(x) (\nabla^\nu R_\nu^{\text{cl}\,I})(x')\, \p^\mu_x F(x,x')\cdot F(x,x').
\end{equation}
As $x\to x'$, the propagator has a logarithmic divergence (in terms of either $\zeta$ or interval in the embedding space, see (\ref{propemb}) below). Hence, the singular part of the integrand $\p_\mu F(x,x') \cdot F(x,x')$ behaves as $\frac{\log |x-x'|}{|x-x'|}$, which gives UV-finite result after the integration over $d^2x$.

Therefore, the expression which contains the divergent contribution is as follows:

\begin{equation}
\delta S^{(2)} = \dfrac{iC_v(G)}{4} \ili d^2x\,d^2x' \sqrt{-g(x)} \sqrt{-g(x')} R_\mu^{\text{cl}\,I}(x) R_\nu^{\text{cl}\,I}(x')\, \p^\mu_x F(x,x') \p^\nu_{x'} F(x,x'),\label{efact}
\end{equation}
where $C_v(G)$ is the value of the Casimir operator in the adjoint representation: $f^{IKL}f^{JKL} = C_v(G)\delta^{IJ}$.

It is possible to calculate the integral in the AdS local coordinates, but we will use another approach to make the AdS covariance manifest. Namely, we move to the embedding $\mathbb{R}^3$ space where we use cartesian coordinates $X^A$ with the metric $\eta_{AB}$ described in subsection (\ref{adsint}). One can restrict integration from $\mathbb{R}^3$ to AdS by introducing the delta--function: $\sqrt{-g}d^2x = 2\delta(X^2-1)d^3X$. The embedding of AdS in $\mathbb{R}^3$ is given by functions $X^A(x) = \varphi^A(x)$ defined in (\ref{embcoord}). Hence, the pushforward of vector fields on AdS to $\mathbb{R}^3$ and the pullback inverse are as follows:

\begin{equation}
R^A(X) = R^\mu(x) \p_\mu \varphi^A(x),\quad R_\mu(x) = R_A(X)\p_\mu \varphi^A(x) .\label{pfrd}
\end{equation}
In terms of AdS vector fields one can write the integrand as $R^{cl}_x(F(x,x'))\cdot R^{cl}_y(F(x,x'))$. In order to use this expression in the embedding space we have to extend the propagator on to $\mathbb{R}^3$. The extension is not unique, but $R^{\text{cl}}$ is tangent to AdS, hence the result does not depend on the choice. 

When $X = \varphi(x)$ and $X' = \varphi(x')$ we have $\zeta+ 1 = \frac{(X+X')^2}{2}$ and $\zeta-1 = -\frac{(X-X')^2}{2}$. We therefore use the following extension for the propagator:
\begin{equation}
F(X,X') = \dfrac{1}{4\pi}\left[ \log \dfrac{1}{-(X-X')^2+i\ep} - \log \dfrac{1}{(X+X')^2+i\ep} \right].
\label{propemb}
\end{equation}
It is convenient to introduce the variables $Y = \frac{X'-X}{2}$ and $\bar{Y} = \frac{X'+X}{2}$. Then, using (\ref{pfrd}), we obtain:

\begin{equation}
\begin{aligned}
\delta S^{(2)} &= \dfrac{iC_v(G)}{4 \pi^2}\ili d^3 Y d^3\bar{Y}\delta((Y+\bar{Y})^2-1)\delta(4Y\bar{Y}) R^{\text{cl}\,A}(\bar{Y}-Y) R^{\text{cl}\,B}(\bar{Y}+Y)\times\\
&\times \left[ -\dfrac{ Y_A Y_B}{\<Y^2-i\ep\>^2} + \dfrac{\bar{Y}_A \bar{Y}_B}{\<\bar{Y}^2+i\ep\>^2} + \dfrac{Y_A\bar{Y}_B + \bar{Y}_AY_B}{(Y^2-i\ep)(\bar{Y}^2+i\lc)} \right].
\end{aligned}
\end{equation}
Here we have transformed the delta-functional term for convenience. 

The first two terms in the last expression are logarithmically divergent and can be calculated in the usual way via Wick rotation. The third term there is convergent and can be omitted. Then the UV--divergent parts of the first and the second terms are as follows:

\begin{equation}
\begin{aligned}
\delta S^{(2)}_{\text{loc}} &= -\dfrac{iC_v(G)}{4\pi^2}\ili d^3Y d^3\bar{Y} \delta(Y^2-1)\delta(4Y\bar{Y}) R^{\text{cl}\,A}(Y)R^{\text{cl}\,B}(Y) \dfrac{\bar{Y}_A\bar{Y}_B}{(\bar{Y}^2-i\ep)^2};\\
\delta S^{(2)}_{\text{non-loc}} &= \dfrac{iC_v(G)}{4\pi^2}\ili d^3Y d^3\bar{Y} \delta(Y^2-1)\delta(4Y\bar{Y}) R^{\text{cl}\,A}(-Y)R^{\text{cl}\,B}(Y) \dfrac{\bar{Y}_A\bar{Y}_B}{(\bar{Y}^2+i\ep)^2},
\end{aligned}
\end{equation}
in the first term we have interchanged the variables $Y$ and $\bar{Y}$. These expressions are manifestly isometry-covariant as the isometries, being Lorentz transformations of the ambient space--time, are coordinate-independent. Due to the symmetry considerations the term $\bar{Y}_A\bar{Y}_B$ can be replaced with $\frac{1}{2} (\eta_{AB} - Y_AY_B)\bar{Y}^2$ as $\bar{Y}$ is orthogonal to $Y$ and $Y^2 = 1$ because of the delta-functions. Furthermore, $R^{\text{cl}\,A}(Y)$ is tangent to AdS, meaning that $R^{\text{cl}\,A}Y_A = 0$ when $Y^2 = 1$. Then only the contribution from $\eta_{AB}$ term survives. What remains to be calculated is the following integral:

\begin{equation}
\mt{I}_{\pm} = \ili d^3\bar{Y} \delta(4Y\bar{Y})\dfrac{\bar{Y}^2}{(\bar{Y}^2\pm i\ep)^2}.
\end{equation}
As $Y^2 = 1$ and the integrand except for the delta-function is Lorentz-invariant, it is possible to set $Y = (0,0,1)$ by Lorentz transformations and integrate out the delta-function. What left is a two-dimensional integral which can be calculated after the Wick rotation $\bar{Y}^1 \to \pm i \bar{Y}^1$. Introducing the UV-cutoff $\frac{1}{\Lambda}$ at small $\bar{Y}$, we obtain:
\begin{equation}
\mt{I}_{\pm} = \mp \dfrac{\pi i}{2} \log \Lambda + \text{UV~finite~terms}.
\end{equation}
Using the pullback (\ref{pfrd}) and the relation $\p_\mu \varphi^A(x) \p_\nu \varphi_A(x) = g_{\mu\nu}(x)$ alongside with the property $\phi(\bar{x}) = -\phi(x)$ of our coordinate choice, we obtain the effective action $\Gamma[\grg^{\text{cl}}]$ up to the second order:
\begin{equation}
\begin{aligned}
\Gamma[\grg^{\text{cl}}] &= -\dfrac{1}{2} \ili d^2x \sqrt{-g}\Tr\left[\<\dfrac{1}{e_0^2}-\dfrac{C_v(G)}{4\pi}\log\Lambda\>R^{\text{cl}}_\mu(x) R^{\text{cl}\,\mu}(x) -\right.\\
&-\left. \dfrac{C_v(G)}{4\pi}\log \Lambda\,\zeta^{\mu\nu}(x,\bar{x})R_\mu(x)R_\nu(\bar{x})\right],\label{eact}
\end{aligned}
\end{equation}
where $\zeta^{\mu\nu}(x,x')$ is a bitensor defined as follows:
\begin{equation}
\zeta^{\mu\nu}(x,x') = \p^\mu \varphi^A(x) \p^\nu \varphi_A(x') = \p^\mu_x \p^\nu_{x'}\zeta(x,x').
\end{equation}
The local term in (\ref{eact}) coincides with the result in flat space.
Apparently the lagrangian (\ref{eact}) is covariant, but it contains the non-local term. Similar term was obtained in the scalar field theory \cite{Akhmedov2018Ultraviolet}. Such a term does not appear if only the local singularity has the Feynman $i\ep$ prescription. That is the case only if the propagator satisfies the equation with only local delta-function on the RHS of such an equation as (\ref{KGeq}). 

\subsection{Effective action from the periodic propagator in global AdS}

\noindent In the case of massive theory the effective action in global AdS can also be defined using the periodic propagator (\ref{temp}). Due to the discussion in the subsection \ref{masst}, in this case the perturbation theory is more justified than in the previous subsection. However, as we will see now the structure of this propagator leads to severe UV divergences. 

To illustrate the origin of such UV divergences let us consider the real scalar $\phi^4$ theory with the mass $m$ and coupling constant $\lambda$. The calculation of the effective action is similar to \cite{Akhmedov2018Ultraviolet}, but with the use of the periodic propagator (\ref{temp}) instead of the AdS time-ordered one. The contribution to the loop correction, which is of interest for us is as follows:

\begin{equation}
\delta S_{WW} \propto i\lambda^2 \ili d^4x\,d^4 x'\, \sqrt{-g(x)}\,\sqrt{-g(x')} \phi^2(x) \phi^2(x') W_m(x,x')W_m(x',x),
\end{equation}
here $W_m(x,x')$ is the Wightman function of the theory. There is a coordinate system where $W_m(x,x')$ asymptotically coincides with the Wightman function in Minkowski space for close $x$ and $x'$. Hence, the qualitative expression for the contribution coming from the local singularity ($x'$ is on a light cone emanating from $x$) is as follows:

\begin{equation}
\delta S_{WW}^{(M)} \propto i\lambda^2 \ili d^4x\,d^4 x' \dfrac{\phi^2(x) \phi^2(x')}{\<(x-x')^2 - i\ep \sign (x^0-x'^0)\>\<(x-x')^2 + i\ep \sign (x^0-x'^0)\>},
\end{equation}
where we used the expression for the Wightman function in Minkowski space--time for close $x$ and $x'$. The leading UV term is mass-independent. The signature of metric is $(+,-,-,-)$. Introducing the coordinates $y = \frac{x'-x}{2}$ and $y' = \frac{x'+x}{2}$, we can rewrite the latter term as:

\begin{equation}
\delta S_{WW}^{(M)} \propto i\lambda^2 \ili d^4y\,d^4 \bar{y} \dfrac{\phi^2 [\bar{y}-y]\phi^2[\bar{y}+y]}{(y^0-|\bd{y}| - i\ep)(y^0+|\bd{y}| - i\ep)(y^0-|\bd{y}| + i\ep)(y^0+|\bd{y}| + i\ep)},
\end{equation}
where $\bd{y} = (y^1,y^2,y^3)$. As one can see there are poles in both upper and lower half-planes of $y^0$. The $y^0$-integration is now divergent when $\ep \to 0$. The most divergent part can be obtained by setting $y^0 = \pm |\bd{y}|$ in the denominator and calculating the $y^0$ integral by closing the contour in upper half-plane:

\begin{equation}
\delta S_{WW}^{(M)} \propto \dfrac{i \lambda^2}{\ep} \ili d^4 y'\,d^3 y \<\dfrac{\phi^2 [y'-y]\phi^2[y'+y]|_{y^0 = |\bd{y}|} + \phi^2 [y'-y]\phi^2[y'+y]|_{y^0 = -|\bd{y}|}}{|\bd{y}|^2}\>. \label{dSWW}
\end{equation}
The remaining integral is convergent.

In this case it is natural to associate $\ep$ with UV-cutoff $\frac{1}{\Lambda}$. Hence, the divergence is powerlike, but it cannot be subtracted by a local counterterm as is usually done with powerlike divergences due to the obvious non-locality in (\ref{dSWW}). This non-locality is concentrated on the light cone emanating from $y'$. \textcolor{black}{Due to quasiperiodicity of Wightman function on AdS, the similar singularity will appear on the light cone emanating from the antipodal point of the source.}

Such a divergence as in (\ref{dSWW}) is drastically different from what one usually obtains in flat space QFT, because it comes from the product of two Wightman functions with different order of the same points, and, hence, with different signs of $i\ep$-prescription. In the usual situation the time ordering of Feynman propagator, $F_m(x,x') = \theta(t-t')W_m(x,x') + \theta(t'-t)W_m(x',x)$, prevents such cases~--- the product of two step-functions is definitely zero.

\section{Loop corrections in the CAdS and PP}\label{cvpnk}

\subsection{Theory in the CAdS space--time}\label{cover}

\noindent In the CAdS space--time the theory is free of non-local counterterms, if one uses the conventional time-ordering and the corresponding $i\epsilon$--prescription \cite{Akhmedov2018Ultraviolet}. However, there still can be problems in loops due to the quasiperiodicity of the propagator, because naively in the CAdS space--time one repeats the same contribution of the global AdS infinite number of times. Note that due to the quasiperiodicity in time in CAdS the Feynman propagator does not decay with the increase of the timelike distance between its two points. 

In this section we will show that such a theory can be properly defined by analytical continuation from the Euclidean theory and hence is free from these problems. However, that demands an unusual $i\epsilon$--shift in the propagators. Furthermore, in general case the theory is not be strictly causal: the commutator function $\bl[\phi(x),\phi(x')]\br$ may be nonzero for space--like separations between $x$ and $x'$ if $|\tau-\tau'| > \pi$.

First, let us discuss the analytic properties of the propagator. We start from the Wightman function $W^{(s)}(x,x')$ of the free massive theory. For simplicity we set $x' = 0$ (we can always do it by CAdS isometries). As we are interested in the behavior of the Wightman function in the complex $\tau$-plane, it is not very convenient to use the expression (\ref{masswig}), because $\zeta$ is periodic in $\tau$. Hence, the expression (\ref{wigseries}) in terms of power series obtained in appendix \ref{NormWig} is more suitable:
\begin{equation}
    W^{(s)}(x,0) = \dfrac{\Gamma(s)}{2\sqrt{\pi}\Gamma(s+1/2)} \cos^{s}\theta e^{-is (\tau-i\ep)} \sum_{k=0}^{\infty} e^{-ik(2\tau - i\ep)} \dfrac{(s)_k}{(s+1/2)_k} P_k^{(-1/2,s-1/2)}(\cos 2\theta).\label{wigseries1}
\end{equation}
We see that this function is clearly analytic in the lower half plane $\text{Im}\, \tau < 0$ as the series is convergent and for real $\tau$ the $i\ep$-prescription yields the lower boundary value. The function $W(0,x)$ can be obtained by simply changing the sing of $\tau$, as the general expression (\ref{wightsum}) is symmetric with respect to the interchange of $\theta$ and $\theta'$. Hence, $W^{(s)}(0,x)$ is analytic in the upper half-plane. 

The Legendre function (\ref{masswig}) has branching points at $\zeta \in \{1,-1,\infty\}$ (for the moment we forget about the $i\lc$--shift), which corresponds to $\cos \tau = \pm \cos \theta$ and $\cos\theta = 0$, i.e. $\tau \in \{\pm \theta + \pi k, \pi/2 + \pi k\}$, $k\in \mathbb{Z}$. In the region $-|\theta| <\tau < |\theta|$ (the separation is spacelike, since $\zeta > 1$) both Wightman functions $W(x,0)$ and $W(0,x)$ coincide and are given by $\frac{1}{2\pi} Q_{s-1}(\zeta)$, as $\zeta > 1$ and the branch cut in $\zeta$ plane is $(-\infty,1]$. Therefore, the commutator function $\bl [\phi(x), \phi(0)]\br$ is zero in this region. 

However, the theory is not completely causal for generic mass or $s$. In fact, while $\zeta$ is periodic and the spacelike regions $|\zeta| > 1$ appear every $\pi$-period, the functions $W^{(s)}(x,0)$ and $W^{(s)}(0,x)$ get multiplied by $e^{- i \pi s}$ and $e^{ i \pi s}$ respectively. They coincide and therefore the causality is satisfied only if $s$ is integer, i.e. when the Wightman function can be correctly defined on global AdS.

\begin{figure}[h]
\centering\includegraphics[width=0.85\textwidth]{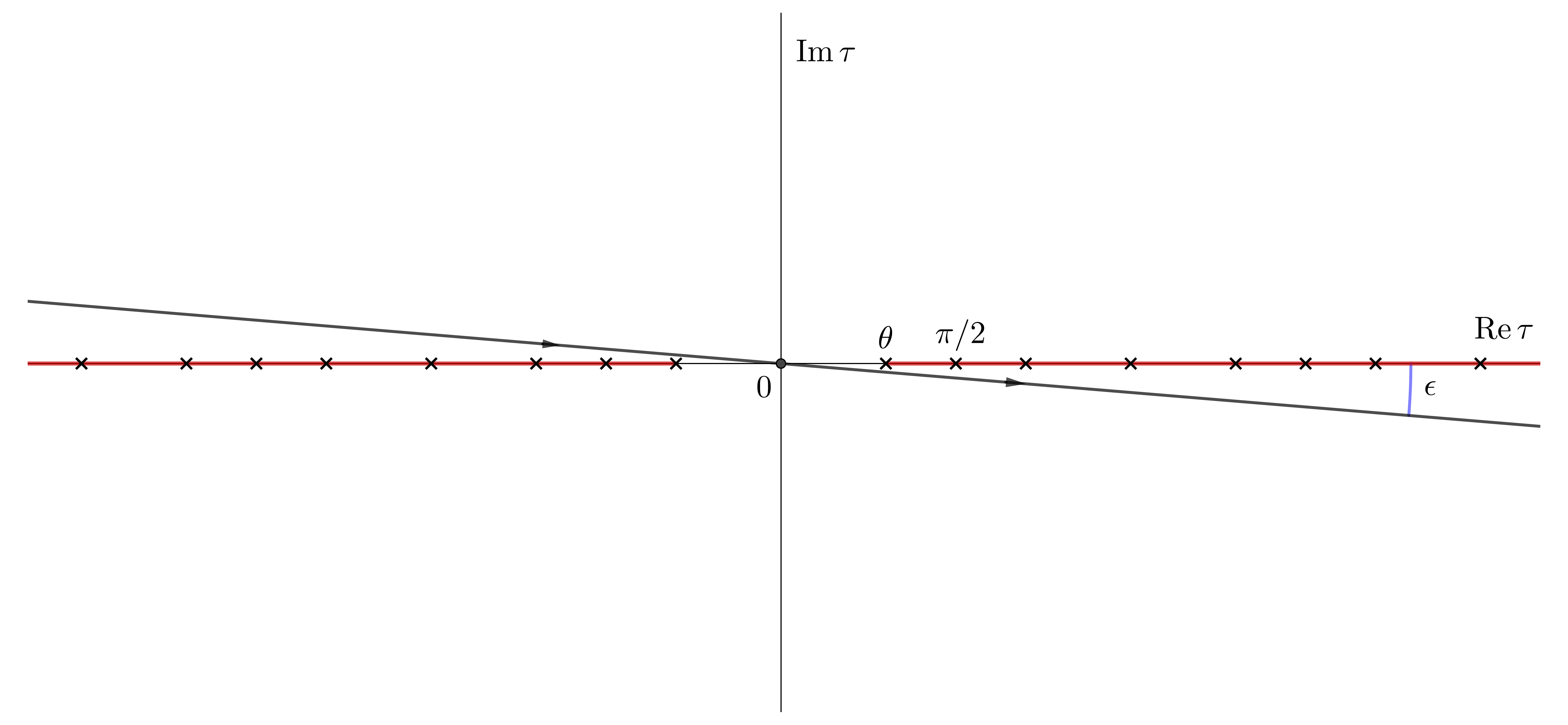}
\caption{The complex $\tau$-plane of the Feynman propagator with the integration contour: branching points are marked by crosses and branch-cuts by red lines. } \label{cmptau}
\end{figure} 

In the same way as it is done in Minkowski space--time quantum field theory the functions $W^{(s)}(x,0)$ and $W^{(s)}(0,x)$  can be glued together into the Feynman propagator $\tilde{F}^{(s)}(x,0)$ through the cut $-|\theta| < \tau < |\theta|$ (the expression is given in (\ref{massprop})). This region is in fact $S_0^+(0)$. The propagator is analytic in $\mathbb{C}\setminus \{(-\infty,-|\theta|] \cup [|\theta|,+\infty)\}$ as a function of $\tau$. If $\tau \in \mathbb{R}$ we choose the lower boundary value for $\tau > 0$ and the upper one for $\tau \le 0$. \textcolor{black}{However, similarly to the discussion in subsection \ref{masst}, the causality in $S_0^+(0)$ is enough for the propagator to be invariant under the isometry group of the CAdS space--time.}

In perturbation theory one has to calculate the integrals involving $\tilde{F}^{(s)}(x,0)$ over the $\tau$-axis. To do that one has to specify the integration contour. The standard contour parallel to the real axis and going slightly above or below (depending on the $\sign \tau$) is not consistent in CAdS for sufficiently large $|\tau|$. In fact, along such a contour the Feynman propagator in CAdS is quasiperiodic and does not decay at infinity. As a result, the loop integrals are ill-defined.  However, such a choice of the contour is the standard one in Minkowski space-time: it is the proper contour to define the time evolution of correlation functions in the Schwinger--Keldysh diagrammatic technique.  

The well defined choice in CAdS is provided by the contour $e^{-i \ep}\mathbb{R}$ for $\pi > \ep > 0$. As one can see, it is quite different from the standard way of doing the $i\epsilon$--shift in Minkowski space--time quantum field theory. For this new type of the contour the correlation function exponentially decays at infinity in CAdS, which can be seen directly from the expression for Wightman function (\ref{wigseries1}) and the fact that $s > 1$ (in the upper half-plane we should change $\tau$ to $-\tau$). An example of such a contour with the branching points and branch cuts in $\tau$-plane is shown on the fig. \ref{cmptau}. 

More importantly, the contour can be deformed into the Euclidean one with $\ep = \pi/2$. After such a Wick rotation $\tau = -i\beta$ the covering space is mapped into the upper half of Euclidean AdS space (EAdS), which is maximally symmetric. In $d$-dimensional case we have:
\begin{equation}
    X^i = \psi^i\tan\theta,\quad X^{d-1} = \dfrac{\cosh\beta}{\cos\theta},\quad \tilde{X}^d = \dfrac{\sinh\beta}{\cos\theta},\quad \zeta(x,0) = \dfrac{\cosh\beta}{\cos\theta} > 1,
\end{equation}
where $X^d = - i \tilde{X}^d$, the Wick-rotated time in the embedding space.

In the Euclidean theory the Feynman propagator $\tilde{F}^{(s)}(x,0)$ is singular only when $x = 0$. Hence all divergences are local and the non--local counterterms, which we have observed in the previous section, do not arise. As the upper half of EAdS is maximally symmetric, the correlation functions and loop corrections depend only on the isometry invariants (there are no problems associated with branch-cuts as $\zeta \ge 1$ for every two points in EAdS). The analytic continuation back to Minkowski space violates this property~--- the $i\ep$-prescription in $\zeta$-plane, which is $\zeta + i\ep \sin |\tau|$ depends on $\tau$. Instead we get the invariance under the isometry of the CAdS space--time. In the next subsection we will show that the theory in the PP can also be Wick-rotated to EAdS theory, but the analytic continuation in this case preserves the invariance~--- the propagator depends only on $\zeta$.



\subsection{Theory in Poincare patch}

\noindent The PP is defined in (\ref{poincare}). Comparing with the global embedding (\ref{embcoord}), we see that standard time--ordering in PP corresponds to AdS--invariant ordering (\ref{tord}) for timelike separated points. Hence, the Feynman propagator in PP has the same $i\ep$-prescription as the AdS time-ordered one in the global chart. The propagator depends only on the isometry invariants, $\zeta = XX'$ and $\sign \Delta t$. It is also straightforward to see that the Wightman's prescription $t-t'-i\ep$ for two-point function $\bl \phi(x) \phi(x') \br$ corresponds to the shift $\zeta(x,x') + i\ep \sign (t-t')$. 

The isometry invariant in this case (\ref{PoincareMetr}) is not periodic in time, unlike the case discussed above --- the dependence is almost the same as in flat Minkowski space. Hence, the simple $i\ep$-shift in the complex $t$--plane is enough to correctly define the theory. Also $\zeta$ does not encircle branching points after any $t$-translations. Hence, the monodromy phase multipliers like in (\ref{massprop}) do not appear. 

The Feynman propagator $\mathfrak{F}(x,x')$ therefore has the following form:

\begin{equation}
\mathfrak{F}(x,x') = \mathfrak{F}(\zeta + i\ep),
\end{equation}
it is the boundary value of a function, regular in the upper half-plane which satisfies the equations of motion. It means that in the case of two-dimensional massive scalar field theory it should coincide with $F^{(s)}(x,x')$ defined in (\ref{massfeyn}).

Let us consider now loop corrections in a quantum field theory in PP of AdS. There are isometry transformations, which move the PP in the global AdS. Hence, naively if one integrates over the PP in the loop integrals the isometry is broken. However, now we will show that loop corrected correlators are still functions of the isometry invariant for the AdS isometry state, i.e. for the Feynman propagator defined above. The arguments in this subsection are very similar to those in \cite{Polyakov:2012uc}, \cite{Akhmedov:2013vka} for the expanding Poincare region of de Sitter space--time with the Bunch--Davies initial state exactly at past infinity.

Consider a vertex of order $n$ in a loop contribution. In terms of the embedding space coordinates with the measure discussed in subsection \ref{EfPCF} the integration in the vertex is as follows:
\begin{equation}
I = \ili d^{d+1} Y \,2\delta(Y^2-1) \, \theta(Y^{d-1}-Y^0) \, \prod_{i=1}^n \mathfrak{F}(Y,X_i).
\end{equation}
The step function $\theta(Y^{d-1}-Y^0)$ restricts the domain of integration to the PP and, hence, naively violates the AdS isometry, while $\delta(Y^2-1)$ and $\prod_{i=1}^n \mathfrak{F}(Y,X_i)$ are invariant. To probe the AdS invariance of loop corrections one should check how $I$ is changed under a transformation which does not preserve the PP, i.e. affects $\theta(Y^{d-1}-Y^0)$. For example consider an infinitesimal boost in $(0,d)$-plane:

\begin{equation}
X^0 \to X^0 + \lc X^{d},\quad X^{d} \to X^{d} + \lc X^0.
\end{equation}
The consideration of other boosts and rotations moving either $X^0$ or $X^{d-1}$ is similar. The corresponding change of $I$ is as follows:

\begin{equation}
\delta_\lc I = -\lc \ili 2 d^{d+1} Y\delta(Y^2-1) \, \delta(Y^{d-1}-Y^0) \, Y^{d}\prod_{i=1}^n \, \mathfrak{F}(Y,X_i).
\end{equation}
Let us take into account that $\mathfrak{F}(Y,X_i) = \mathfrak{F}(YX_i+i\ep)$ and

\begin{equation}
Y X = \dfrac{1}{2}\Big[X^{d-1}+X^0)(Y^{d-1}-Y^0)+(X^{d-1}-X^0)(Y^{d-1}+Y^0)\Big] + X^{d}Y^{d} - X^iY^i,~i = 1,\dots,d-2.
\end{equation}
Therefore, if all points $X_i$ are in PP, after integrating out $\delta(Y^{d-1}-Y^0)$ using $d Y^{d-1} d Y^0 = \frac{1}{2} \, d(Y^{d-1}+Y^0) \, d(Y^{d-1} - Y^0)$ we find that:

\begin{equation}\label{5.8}
\begin{aligned}
\delta_\lc I &= -\lc \ili  d(Y^{d-1}+Y^0) \, d^{d-1} Y \,  \delta(Y^2-1) \, Y^{d-1} \times\\
&\times \prod_{j=1}^n \mathfrak{F}\left[\dfrac{1}{2}(X^{d-1}_j-X^0_j)(Y^{d-1}+Y^0+i\ep)+X_j^{d}Y^{d}-X_j^iY^i\right],\quad i = 1,\dots,d-2.
\end{aligned}
\end{equation}
Here we have used that $X^{d-1}-X^0 > 0$ in the PP. In (\ref{5.8}) we can close the integration contour in upper half of complex $(Y^{d-1}+Y^0)$--plane where the integrand has no poles. As the result we obtain that $\delta_\lc I = 0$ and the integral $I$ (and hence the loop correction) does not change under isometry transformations. Hence, $I$ depends only on isometry invariants, such as $\zeta$ and sign of $\Delta t$. This is in agreement with the possibility of the Wick rotation, which we will discuss now.

The theory in PP also admits the Wick rotation $t \to -it$. It is equivalent to the transformation $X^d \to -i X^d$ in terms of embedding space, which brings the PP to the upper sheet of the two-sheeted EAdS space. On the other hand, the Wick rotation of the theory in the CAdS space--time discussed in subsection \ref{cover} transforms it into the very same sheet of EAdS. However, the embedding coordinates transform differently~--- it leads to a different choice of Euclidean time. As we have discussed above, the upper half of EAdS is maximally symmetric. Hence, the correlation function of isometry--invariant Euclidean theory (along with loop integrals) depends only on $\zeta$. The analytical continuation back to PP requires the simple shift $\zeta \to \zeta+ i\ep$. Hence, it preserves the invariance. Also the countertems are inherited from the Euclidean theory and therefore local.

Another consequence of the possibility of Wick rotation is that the propagators on the CAdS and the PP are analytic continuations of the same function of $\zeta$, but with different choices of the complex parameter. As a result, the only difference between them is $i\ep$-prescription~--- on the CAdS space its sign depends on time, which leads to complicated behaviour. Moreover, the Feynman propagator on the CAdS space, $\tilde{F}(x,0)$, is equal to $\mathfrak{F}[\zeta(x,0) + i\ep]$ if $\tau \in (-\pi,\pi)$. It is so because $\tau = 0$ also belongs to the Euclidean contour. Hence, the function at this point is the same as the Euclidean one. Next,  the region including $\tau = 0$ where $\zeta$-plane $i\ep$-prescription on the covering space ($\zeta + i\ep \sin|\tau|$) coincides with one on the PP is $[-\pi,\pi]$~--- we observed it in case of $d=2$ massive free theory. If we increase $\tau$ further so that $i\ep$ prescription is again the same (e.g. $\tau \in [2\pi,3\pi]$), $\zeta$ encircles the branching points, so nontrivial monodromy may change its value.   

\section{Conclusion} 

We discuss quantum field theory in asymptotically AdS space--times, covering of global AdS manifold and in Poincare patch. The relation of the presence of non-local UV singularities of the Wightman function to the classical geodesic focusing is investigated; we find that the latter takes place only in specific cases, while the UV singularity emanating from the antipodal point of the source appears for a wider class of asymptotically AdS metrics. 

Then we consider interacting field theories in 2D global AdS, its covering space and Poincare region. We consider different possible definitions of the Feynman propagator with the focus on the interplay between the time--ordering, $i\epsilon$--shift and isometry invariance. For isometry-invariant propagator we find that perturbation theory becomes inconsistent or leads to the appearance of non-local counterterms in global AdS space--time. It is demonstrated on the example of principal chiral field. We also show that such problems are absent for the theory in Poincare patch. In the latter case quantum field theory over the AdS invariant state can be defined via analytical continuation from the Euclidean AdS space.

In the covering global AdS space--time the analytical continuation can also be done for a specific choice of the $i\epsilon$--shift of the UV singularity in coordinate space. In such a case one can use the isometry--invariant propagator and can Wick--rotate to the Euclidean AdS. But the $i\epsilon$--shift in question is different from the standard one used to define Feynman propagator in flat space--time. 

Let us clarify the last point. If one were considering a non--Planckian initial distribution in covering AdS space--time, one would use Schwinger--Keldysh diagrammatic technique rather than the Feynman one to calculate the time evolution of the distribution. In that technique there is the matrix of propagators one component of which is the Feynman propagator. To define the latter propagator one cannot use the $i\epsilon$--shift adopted in our paper.

\section{Acknowledgements}

We would like to acknowledge valuable discussions with T.Damur, U.Moschella, F. Popov and A.Polyakov.

The work of ETA was supported by the grant from the Foundation for the Advancement of Theoretical Physics and Mathematics ``BASIS'' and by RFBR grant 18-01-00460. The work of ETA and IVK is supported by Russian Ministry of education and science (project 5-100). The work of AAA and IVK was supported by the Russian Science Foundation grant (project no. 18-12-00439).

\begin{appendices}
\section{Boundary conditions and conservation laws}\label{massl}
\noindent We consider the 2-dimensional theory in global AdS space--time with the following action:
\begin{equation}
S = \ili d^2x \sqrt{-g} \<\dfrac{1}{2}(\p_\mu \phi)^2 - V(\phi)\>.
\end{equation}
Following \cite{Avis1978AdS}, we impose the condition of the conservation of charges corresponding to the conserved currents. Namely, $\text{AdS}_2$ has 3-dimensional isometry group $SO(2,1)$ generated by three Killing vectors $K_{AB}^{\mu}$, $A < B$. The covariantly conserved currents can be expressed as follows:
\begin{equation}
J_{AB}^{\mu} = T^\mu_\nu K_{AB}^{\nu},\label{curr}
\end{equation}
where $T_{\mu\nu}$ is a stress-energy tensor of the theory:
\begin{equation}
T_{\mu\nu} = \p_\mu \phi \p_\nu \phi - \dfrac{1}{2} g_{\mu\nu} (\p_\lambda \phi)^2.
\end{equation}
As there are conservation laws of the form $\p_\mu (\sqrt{-g} J_{AB}^{ \mu}) = 0$, for each current there is a corresponding charge:
\begin{equation}
Q_{AB} = \ili_{\tau = \const}d\theta\, \sqrt{-g}J_{AB}^{0} = \ili_{\tau = \const} d\theta\, \sqrt{-g} g^{00} T_{0\mu} K_{AB}^{ \mu}.\label{charge}
\end{equation}
Conservation of these charges means that there is no flux of $\sqrt{-g} J_{AB}^{ \mu}$ through the boundaries $\theta = \pm \pi/2$:
\begin{equation}
(\sqrt{-g}g^{11}T_{1\mu} K_{AB}^{\mu})|_{\theta = \frac{\pi}{2}} - (\sqrt{-g}g^{11}T_{1\mu} K_{AB}^{\mu})|_{\theta = -\frac{\pi}{2}} = 0.\label{conscond}
\end{equation}
Note that this condition is Weyl-invariant.

The isometries of AdS correspond to Lorentz transformations of the embedding space (\ref{embcoord}) generated by the Killing vectors $\mt{K}_{AB} = X_A \p_B - X_B \p_A$. They are tangent to AdS, hence $K_{AB}$ are their restrictions to $X^2 = 1$. In terms of local coordinates $(\tau,\theta)$ we find:
\begin{equation}
\begin{aligned}
K_{01} &= \sin\tau \sin\theta \p_\tau - \cos\tau\cos\theta \p_\theta;\\
K_{12} &= \p_\tau;\\
K_{02} &= -\cos\tau \sin\theta \p_\tau - \sin\tau \cos\theta \p_\theta.
\end{aligned}\label{killing}
\end{equation}
Substituting these explicit expressions into (\ref{conscond}), we obtain:
\begin{equation}
T_{10}|_{\theta = \frac{\pi}{2}} = T_{10}|_{\theta = -\frac{\pi}{2}} = 0.\label{bcond}
\end{equation}
As $T_{10} = \p_1\phi \p_0\phi$, at $\theta = \pm \pi/2$ either $\p_0 \phi = 0$ or $\p_1 \phi = 0$. Hence, there are four types of boundary condition. In terms of modes the first condition means that the modes are zero at the boundary (as they are eigenfunctions of $\p_0$), and the second condition means that their derivatives are zero. If we additionally impose the condition of maximal decay of Wightman function at infinity, we have to choose the first one. Therefore,
\begin{equation}
u_n(\pi/2) = u_n(-\pi/2) = 0,\label{bcondr}
\end{equation}
$u_n$ are defined in (\ref{phiop}).

\section{Normalization of modes and Wightman function of massive theory}\label{NormWig}
\noindent Let us denote $v_n = u_{2n-1}$, $w_n = u_{2n}$ from (\ref{massspectr}). To find the normalization constants we have to calculate the scalar products and use the condition (\ref{norm}):
\begin{equation}
\begin{aligned}
\braket{v_n}{v_n} &= A_n^2 \ili d\theta \cos^{2s}\theta\<P_{n-1}^{(-1/2,\,s-1/2)}(\cos 2\theta)\>^2 = \\ &= \frac{ A_n^2}{2^{s}} \ili_{-1}^1 dx\, (1+x)^{s-1/2} (1-x)^{-1/2}\<P_{n-1}^{(-1/2,\,s-1/2)}(x)\>^2 =\\
&= \dfrac{A_n^2}{2^s} \dfrac{2^s\Gamma\<n-1/2\> \Gamma\<n+s-1/2\>}{(2(n-1)+s) \Gamma(n+s-1)\Gamma(n)},
\end{aligned}
\end{equation}
here we used the orthogonality relation for Jacobi polynomials and the change of the integration variables $\cos 2\theta = x$. Therefore,
\begin{equation}
A_n = \sqrt{\dfrac{\Gamma(n+s)(n-1)!}{2\Gamma(n-1/2)\Gamma(n+s-1/2)}}.
\end{equation}
Next,
\begin{equation}
\begin{aligned}
\braket{w_n}{w_n} &=  B_n^2  \ili d\theta \cos^{2s} \theta \sin^2 \theta \<P_{n-1}^{(1/2,\,s-1/2)}(\cos 2\theta)\>^2 =\\
&= \dfrac{B_n^2 }{2^{s+1}} \ili_{-1}^1 dx(1+x)^{s+1/2}(1-x)^{1/2} \<P_{n-1}^{(1/2,\,s-1/2)}(x)\>^2= \\
&= \dfrac{B_n^2}{2^{s+1}} \dfrac{2^{s+1}\Gamma(n+1/2)\Gamma(n+s-1/2)}{(2n+s-1)\Gamma(n+s)\Gamma(n)}.
\end{aligned}
\end{equation}
And, hence:
\begin{equation}
B_n = \sqrt{\dfrac{\Gamma(n+s)n!}{2\Gamma(n+1/2)\Gamma(n+s-1/2)}}.
\end{equation}
Now we can calculate the Wightman function using (\ref{wightsum}):
\begin{equation}
W(x,x') = \sum_{n=1}^{\infty}\<e^{i(s+2n-2)(\tau'-\tau +i\ep)} v_n (\theta) v_n(\theta') + e^{i(s+2n-1)(\tau'-\tau +i\ep)}  w_n (\theta) w_n(\theta')\>
\end{equation}
As the answer is AdS invariant, we can set $\theta' = 0,~\tau' = 0$. Then there will be no contribution from $w_n$ as $w_n(0) = 0$. Using
\begin{equation}
P_n^{(\alpha,\beta)}(1) = \dfrac{\Gamma(\alpha+1+n)}{\Gamma(\alpha+1)n!},
\end{equation}
we obtain:
\begin{equation}
\begin{aligned}
W^{(s)}(x,x') &= \sum_{n=1}^{\infty} e^{-i(s+2n-2)(\tau - i\ep)} \cos^{s} \theta \dfrac{\Gamma(n+s-1)}{2\sqrt{\pi} \Gamma(n+s-1/2)} P_{n-1}^{(-1/2,s-1/2)}(\cos 2\theta) = \\
&= \dfrac{\Gamma(s)}{2\sqrt{\pi}\Gamma(s+1/2)} \cos^{s}\theta e^{-is (\tau-i\ep)} \sum_{k=0}^{\infty} \<e^{-i(2\tau - i\ep)}\>^k \dfrac{(s)_k}{(s+1/2)_k} P_k^{(-1/2,s-1/2)}(\cos 2\theta) = \\
&=\dfrac{\Gamma(s)}{2\sqrt{\pi}\Gamma(s+1/2)} \dfrac{e^{-is(\tau-i\ep)} \cos^{s}\theta}{\<1+e^{-i(2\tau-i\ep)}\>^{s}}~_2F_1\<\dfrac{s}{2},\dfrac{s+1}{2};s+\dfrac{1}{2}; \dfrac{2e^{-i(2\tau-i\ep)}(\cos 2\theta + 1)}{\<1+ e^{-i(2\tau-i\ep)}\>^2}\>,\label{wigseries}
\end{aligned}
\end{equation}
where $(x)_n$ is a Pochhammer symbol:
\begin{equation}
(x)_n = \prod_{k=0}^{n-1} (x+k).
\end{equation}
The answer for the infinite sum is given in \cite{Prudnikov2002Series}. In this case the expression for geodesic parameter is as follows:
\begin{equation}
\zeta = f_A(x) f^A(x') = \dfrac{\cos \tau}{\cos \theta},
\end{equation}
Using the standard expression for Legendre function \cite{Bateman1953Transc}, we find:
\begin{equation}
 W^{(s)}(x,x') = \dfrac{1}{2\pi} Q_{s-1}\big[\zeta + i\ep \sin(\tau-\tau')\big].
\end{equation}
That is the propagator used in the main body of the note.

\section{The WKB correction to frequencies}{\label{WKBcor}}
\noindent According to (\ref{D}), we have to calculate the non-vanishing part of the following expression when $\w_n \to \infty$:
\begin{equation}
D_n = 2\pi \w_n - \oint\sqrt{\w_n^2 - f(\theta)}d\theta = 4\<\dfrac{\pi \w_n}{2} - \ili_0^{\theta_t}\sqrt{\w_n^2 - f(\theta)}\>\label{D_n}.
\end{equation}
As the integrand significantly differs from $\w_n$ only near the boundary, we can use the asymptotic expression (\ref{f}). Also, changing the lower limit of integration in (\ref{D_n}) to $-L$ ($L>0$) can be compensated by the term $L \w_n$ as the additional corrections to this compensation vanish when $\w_n \to \infty$. Hence,

\begin{equation}
D_n \approx 4 \lim_{L\to \infty}\<\w_n\<L + \dfrac{\pi}{2} \>  - \ili_{-L}^{\theta_t}d\theta \sqrt{\w_n^2 - \dfrac{C}{(\pi/2-\theta)^{\alpha}}}\> = 4\lim_{L\to \infty}\<\w_n L - \ili_{\tilde{\theta}_t}^{L}d\theta\sqrt{\w_n^2 - \dfrac{C}{\theta^\alpha}}\>,
\end{equation}
where $\tilde{\theta}_t = \<\frac{C}{\w_n^2}\>^{1/\alpha}$, $\alpha \ge 2$. One can show that the limit is finite. Let $L = \tilde{\theta}_t a$, then

\begin{equation}
D_n \approx 4 \lim_{a\to\infty} C^{\frac{1}{\alpha}}\w_n^{1-\frac{2}{\alpha}}\<a- \ili_1^a d\theta \sqrt{1-\dfrac{1}{\theta^\alpha}}\>.
\end{equation}
Let us calculate this integral:
\begin{equation}
I(a) = \ili_1^{a} d\theta \sqrt{1-\dfrac{1}{\theta^\alpha}} = /x = \frac{1}{\theta^\alpha}/ = \dfrac{1}{\alpha}\ili_{a^{-\alpha}}^{1} dx\, x^{-1-\frac{1}{\alpha}} \sqrt{1-x}.
\end{equation}
Integrating by parts, one obtains:
\begin{equation}
I(a) = a - \dfrac{1}{2}\ili_{a^{-\alpha}}^1 \dfrac{x^{-\frac{1}{\alpha}}}{\sqrt{1-x}}.
\end{equation}
In the limit $a\to \infty$ the second term here turns into the Euler's beta function. Therefore,
\begin{equation}
D_n \approx 2 C^{\frac{1}{\alpha}} \w_n^{1-\frac{2}{\alpha}} \ili_0^1 dx \dfrac{x^{-\frac{1}{\alpha}}}{\sqrt{1-x}} = 2 \sqrt{\pi} C^{\frac{1}{\alpha}} \w_n^{1-\frac{2}{\alpha}} \dfrac{\Gamma\<1-\frac{1}{\alpha}\>}{\Gamma\<\frac{3}{2}-\frac{1}{\alpha}\>}.
\end{equation}
As according to (\ref{D_n}) and (\ref{D}) $D_n \approx D\w_n^{1-\frac{2}{\alpha}}$, and we find that:
\begin{equation}
D = 2\sqrt{\pi} C^{\frac{1}{\alpha}} \dfrac{\Gamma\<1-\frac{1}{\alpha}\>}{\Gamma\<\frac{3}{2}-\frac{1}{\alpha}\>},
\end{equation}
which is used in the main body of the note.
\end{appendices}

\printbibliography

@article{Maldacena1999largeN,
title = "Large N field theories, string theory and gravity",
journal = "Physics Reports",
volume = "323",
number = "3",
pages = "183 - 386",
year = "2000",
issn = "0370-1573",
doi = "https://doi.org/10.1016/S0370-1573(99)00083-6",
author = "Ofer Aharony and Steven S. Gubser and Juan Maldacena and Hirosi Ooguri and Yaron Oz",
}

@article{Avis1978AdS,
  title = {Quantum field theory in anti-de Sitter space-time},
  author = {Avis, S. J. and Isham, C. J. and Storey, D.},
  journal = {Phys. Rev. D},
  volume = {18},
  issue = {10},
  pages = {3565--3576},
  numpages = {0},
  year = {1978},
  month = {Nov},
  publisher = {American Physical Society},
  doi = {10.1103/PhysRevD.18.3565},
  url = {https://link.aps.org/doi/10.1103/PhysRevD.18.3565}
}

@article{Akhmedov:2012hk,
    author = "Akhmedov, E.T. and Sadofyev, A.V.",
    title = "{Comparative study of loop contributions in AdS and dS}",
    eprint = "1201.3471",
    archivePrefix = "arXiv",
    primaryClass = "hep-th",
    reportNumber = "ITEP-TH-2-12, AEI-2012-004",
    doi = "10.1016/j.physletb.2012.04.061",
    journal = "Phys. Lett. B",
    volume = "712",
    pages = "138--142",
    year = "2012"
}

@article{Bertan:2018khc,
    author = "Bertan, Igor and Sachs, Ivo",
    title = "{Loops in Anti\textendash{}de Sitter Space}",
    eprint = "1804.01880",
    archivePrefix = "arXiv",
    primaryClass = "hep-th",
    reportNumber = "LMU-ASC 16/18, LMU-ASC-16-18",
    doi = "10.1103/PhysRevLett.121.101601",
    journal = "Phys. Rev. Lett.",
    volume = "121",
    number = "10",
    pages = "101601",
    year = "2018"
}

@article{Carmi:2019ocp,
    author = "Carmi, Dean",
    title = "{Loops in AdS: from the spectral representation to position space}",
    eprint = "1910.14340",
    archivePrefix = "arXiv",
    primaryClass = "hep-th",
    doi = "10.1007/JHEP06(2020)049",
    journal = "JHEP",
    volume = "06",
    pages = "049",
    year = "2020"
}

@article{Ponomarev:2019ltz,
    author = "Ponomarev, Dmitry and Sezgin, Ergin and Skvortsov, Evgeny",
    title = "{On one loop corrections in higher spin gravity}",
    eprint = "1904.01042",
    archivePrefix = "arXiv",
    primaryClass = "hep-th",
    doi = "10.1007/JHEP11(2019)138",
    journal = "JHEP",
    volume = "11",
    pages = "138",
    year = "2019"
}

@article{Bertan:2018afl,
      author         = "Bertan, Igor and Sachs, Ivo and Skvortsov, Evgeny D.",
      title          = "{Quantum $\phi^4$ Theory in AdS${}_4$ and its CFT Dual}",
      journal        = "JHEP",
      volume         = "02",
      year           = "2019",
      pages          = "099",
      doi            = "10.1007/JHEP02(2019)099",
      eprint         = "1810.00907",
      archivePrefix  = "arXiv",
      primaryClass   = "hep-th",
      reportNumber   = "LMU-ASC 63/18",
      SLACcitation   = "%%CITATION = ARXIV:1810.00907;%%"
}

@article{Sleight:2017cax,
    author = "Sleight, Charlotte and Taronna, Massimo",
    title = "{Feynman rules for higher-spin gauge fields on AdS$_{d+1}$}",
    eprint = "1708.08668",
    archivePrefix = "arXiv",
    primaryClass = "hep-th",
    doi = "10.1007/JHEP01(2018)060",
    journal = "JHEP",
    volume = "01",
    pages = "060",
    year = "2018"
}

@article{Bros:2001tk,
    author = "Bros, Jacques and Epstein, Henri and Moschella, Ugo",
    title = "{Towards a general theory of quantized fields on the anti-de Sitter space-time}",
    eprint = "hep-th/0111255",
    archivePrefix = "arXiv",
    doi = "10.1007/s00220-002-0726-z",
    journal = "Commun. Math. Phys.",
    volume = "231",
    pages = "481--528",
    year = "2002"
}

@article{Bros:2011vh,
    author = "Bros, Jacques and Epstein, Henri and Gaudin, Michel and Moschella, Ugo and Pasquier, Vincent",
    title = "{Anti de Sitter quantum field theory and a new class of hypergeometric identities}",
    eprint = "1107.5161",
    archivePrefix = "arXiv",
    primaryClass = "hep-th",
    doi = "10.1007/s00220-011-1372-0",
    journal = "Commun. Math. Phys.",
    volume = "309",
    pages = "255--291",
    year = "2012"
}

@article{Moschella:2007zza,
    author = "Moschella, Ugo and Schaeffer, Richard",
    title = "{Quantum theory on Lobatchevski spaces}",
    eprint = "0709.2795",
    archivePrefix = "arXiv",
    primaryClass = "hep-th",
    doi = "10.1088/0264-9381/24/14/003",
    journal = "Class. Quant. Grav.",
    volume = "24",
    pages = "3571--3602",
    year = "2007"
}

@article{Moschella:2006pkh,
    author = "Moschella, Ugo",
    editor = "Damour, Thibault and Darrigol, Olivier and Duplantier, Bertrand and Rivasseau, Vincent",
    title = "{The de Sitter and anti-de Sitter Sightseeing Tour}",
    doi = "10.1007/3-7643-7436-5_4",
    journal = "Prog. Math. Phys.",
    volume = "47",
    pages = "120--133",
    year = "2006"
}

@article{Polyakov:2012uc,
    author = "Polyakov, A.M.",
    title = "{Infrared instability of the de Sitter space}",
    eprint = "1209.4135",
    archivePrefix = "arXiv",
    primaryClass = "hep-th",
    month = "9",
    year = "2012"
}

@book{Akhmedova:2019bau,
    author = "Akhmedova, Valeriya and Akhmedov, Emil T.",
    title = "{Selected Special Functions for Fundamental Physics}",
    doi = "10.1007/978-3-030-35089-5",
    publisher = "Springer",
    series = "SpringerBriefs in Physics",
    year = "2019"
}

@article{Akhmedov:2013vka,
    author = "Akhmedov, E.T.",
    title = "{Lecture notes on interacting quantum fields in de Sitter space}",
    eprint = "1309.2557",
    archivePrefix = "arXiv",
    primaryClass = "hep-th",
    reportNumber = "ITEP-TH-32-13",
    doi = "10.1142/S0218271814300018",
    journal = "Int. J. Mod. Phys. D",
    volume = "23",
    pages = "1430001",
    year = "2014"
}

@article{Giombi:2017hpr,
    author = "Giombi, Simone and Sleight, Charlotte and Taronna, Massimo",
    title = "{Spinning AdS Loop Diagrams: Two Point Functions}",
    eprint = "1708.08404",
    archivePrefix = "arXiv",
    primaryClass = "hep-th",
    reportNumber = "PUPT-2540",
    doi = "10.1007/JHEP06(2018)030",
    journal = "JHEP",
    volume = "06",
    pages = "030",
    year = "2018"
}

@article{Burgess1985Propagators,
title = "Propagators and effective potentials in anti-de Sitter space",
journal = "Physics Letters B",
volume = "153",
number = "3",
pages = "137 - 141",
year = "1985",
doi = "10.1016/0370-2693(85)91415-7",
author = "C.P. Burgess and C.A. Lütken",
}

@article{Akhmedov2018Ultraviolet,
    author = "Akhmedov, Emil T. and Moschella, Ugo and Popov, Fedor K.",
    title = "{Ultraviolet phenomena in AdS self-interacting quantum field theory}",
    eprint = "1802.02955",
    archivePrefix = "arXiv",
    primaryClass = "hep-th",
    doi = "10.1007/JHEP03(2018)183",
    journal = "JHEP",
    volume = "03",
    pages = "183",
    year = "2018"
}

@article{Castell1968Causality,
    author = "Castell, L.",
    title = "{Causality and conformal invariance}",
    doi = "10.1016/0550-3213(68)90279-4",
    journal = "Nucl. Phys. B",
    volume = "5",
    pages = "601--605",
    year = "1968",
}

@inproceedings{Freiling2001InverseSP,
  title={Inverse Sturm-Liouville problems and their applications},
  author={G. Freiling and V. Yurko},
  year={2001}
}

@book{Birrell1982Quantum,
    author = "Birrell, N.D. and Davies, P.C.W.",
    title = "{Quantum Fields in Curved Space}",
    doi = "10.1017/CBO9780511622632",
    isbn = "978-0-521-27858-4",
    publisher = "Cambridge Univ. Press",
    address = "Cambridge, UK",
    series = "Cambridge Monographs on Mathematical Physics",
    month = "2",
    year = "1984"
}

@book{Maslov1965Asymptotic,
    author = "Maslov, V.P.",
    title = "Perturbation Theory and Asymptotic Methods",
    publisher = "Moscow Univ. Publ.",
    address = "Moscow",
    year = "1965"
}

@book{Landau1976Mechanics,
  author = {Landau, L. D. and Lifshitz, E. M.},
  day = 15,
  edition = 3,
  howpublished = {Paperback},
  isbn = {0750628960},
  month = jan,
  publisher = {Butterworth-Heinemann},
  title = {Mechanics, Third Edition: Volume 1 (Course of Theoretical Physics)},
  url = {http://www.worldcat.org/isbn/0750628960},
  year = 1976
}

@book{Prudnikov2002Series,
author = {Prudnikov, A.P. and Brychkov, Yury and Marichev, O.I.},
year = {2002},
publisher = {Taylor \& Francis},
address = "London, UK",
title = {Integrals and Series. Volume 2, Special functions}
}

@book{Bateman1953Transc,
      author        = "Bateman, Harry and Erdélyi, Arthur",
      title         = "{Higher transcendental functions}",
      publisher     = "McGraw-Hill",
      address       = "New York, NY",
      series        = "Bateman Manuscript Project",
      year          = "1953",
}

@book{Landau1981Quantum,
  author = {Landau, L. D. and Lifshitz, L. M.},
  edition = 3,
  howpublished = {Paperback},
  isbn = {0750635398},
  month = jan,
  publisher = {Butterworth-Heinemann},
  title = {Quantum Mechanics Non-Relativistic Theory, Third Edition: Volume 3},
  url = {http://www.worldcat.org/isbn/0750635398},
  year = 1981
}
\end{document}